\documentclass[journal]{IEEEtran}
\IEEEoverridecommandlockouts
\usepackage{bbm}
\usepackage{times}
\usepackage[cmex10]{amsmath}
\usepackage{amssymb}
\usepackage{epsfig,verbatim}
\usepackage{algorithm,algpseudocode}
\usepackage{hhline}
\usepackage{algpseudocode}
\usepackage{booktabs}
\usepackage{multirow}
\usepackage{graphicx, subfigure}
\DeclareGraphicsExtensions{.pdf,.png,.jpg}
\usepackage{cite}
\usepackage{mathtools}
\usepackage{epsfig,latexsym,cite}
\usepackage{setspace}
\usepackage{multicol}
\usepackage{color}
\usepackage{cuted}
\usepackage{stfloats}

\def\BigRoman{\uppercase\expandafter{\romannumeral\number\count 255 }}
\def\Romannumeral{\afterassignment\BigRoman\count255=}

\long\def\comment#1{}

\newfont{\bbb}{msbm10 scaled 700}

\newfont{\bb}{msbm10 scaled 1100}
\newcommand{\CC}{\mbox{\bb C}}

\newcommand{\av}{{\bf a}}
\newcommand{\bv}{{\bf b}}

\newcommand{\hv}{{\bf h}}

\newcommand{\mv}{{\bf m}}

\newcommand{\rv}{{\bf r}}
\newcommand{\sv}{{\bf s}}

\newcommand{\uv}{{\bf u}}
\newcommand{\wv}{{\bf w}}
\newcommand{\vv}{{\bf v}}
\newcommand{\xv}{{\bf x}}
\newcommand{\yv}{{\bf y}}
\newcommand{\zv}{{\bf z}}

\newcommand{\Am}{{\bf A}}

\newcommand{\Em}{{\bf E}}

\newcommand{\Hm}{{\bf H}}
\newcommand{\Id}{{\bf I}}

\newcommand{\Mm}{{\bf M}}

\newcommand{\Cc}{{\cal C}}

\newcommand{\Pc}{{\cal P}}

\newcommand{\alphav}{\hbox{\boldmath$\alpha$}}

\newcommand{\SNR}{{\sf SNR}}

\renewcommand{\Re}{{\rm Re}}
\renewcommand{\Im}{{\rm Im}}
\newcommand{\eqdef}{\stackrel{\Delta}{=}}

\newcommand{\transp}{{\sf T}}

\newcommand{\BLUE}{\color[rgb]{0,0,0.90}}

\newtheorem{definition}{Definition}

\newtheorem{lemma}{Lemma}

\newtheorem{remark}{Remark}

\newcommand{\argmax}{\operatornamewithlimits{argmax}}

\newtheorem{example}{Example}

\title{Construction of 1-Bit Transmit Signal Vectors for Downlink MU-MISO Systems: QAM constellations}

\author{Sungyeal Park,~\IEEEmembership{Student,~IEEE,} Yunseong Cho,~\IEEEmembership{Student,~IEEE,}
        and~Songnam Hong,~\IEEEmembership{Member,~IEEE}
\thanks{S. Park is with Artificial Intelligence Convergence Network, Ajou University, Suwon, Korea (e-mail: awdrg1541@ajou.ac.kr)}
\thanks{Y. Cho is with the Department of Electrical and Computer Engineering, The University of Texas Austin, TX, USA (e-mail: yscho@utexas.edu)}
\thanks{S. Hong is with the Department of Electronic Engineering, Hanyang University, Seoul,  Korea (e-mail:snhong@hanyang.ac.kr)}        
}

\begin{document}
\maketitle


\begin{abstract}
In this paper, we investigate the construction of a transmit signal for a base station with the massive number of antenna arrays under the cost-effective 1-bit digital-to-analog converters. Due to the coarse nonlinear property, conventional precoding methods could not yield an attractive performance with a severe error-floor problem. Moreover, finding an optimal transmit signal is computationally implausible because of its combinatorial nature. Thus, it is still an open problem to construct a 1-bit transmit signal efficiently. As an extension of our earlier work, we propose an efficient method to construct an 1-bit transmit-signal under quadrature-amplitude-modulation constellations. Toward this, we first derive the so-called feasibility condition which ensures that every user's noiseless observation belongs to a desired decision region, and then transform it as linear constraints. Taking into account the robustness to an additive noise, the proposed construction method is formulated as a well-defined mixed-integer-linear-programming problem. Based on this, we develop a low-complexity algorithm to solve it (equivalently, to generate a 1-bit transmit signal). Via simulations, we verify the superiority of the proposed method in terms of a computational complexity and detection performance.

\end{abstract}


\section{Introduction}\label{sec:intro}

In recent years, massive multiple-input single-output (MISO) has been actively investigated for fifth-generation (5G) and future wireless communication systems due to its significant gain in spectral efficiency \cite{marzetta2010noncooperative}. Because of the large number of antennas, whereas, dealing with a high hardware cost and considerable power consumption become one of the key challenges. In massive MISO systems, the use of cheap and efficient building block, e.g., digital-to-analog converters (DACs) or analog-to-digital converters (ADCs), has attracted the most interest as a promising low-power solution \cite{spencer2004introduction,larsson2014massive}. Considering the same clock frequency and resolution,  it is known that DACs have lower power consumption than ADCs, therefore research on low-resolution DACs are often ignored for this reason.
However, in downlink multiuser massive MISO systems, the number of transmit antenna at base station (BS) is much larger than the number of receive antennas. In this context, we should consider DACs' power consumption, cost, and computation complexity. 
In downlink systems, conventional precoding method such as zero-forcing (ZF) and regularized ZF (RZF) achieve almost optimal performance effectively \cite{peel2005vector}. These linear precoding schemes have a low complexity and widely used in wireless communication with high resolution DACs (e.g., 12 bits). But in reality, massive MISO must be built with low cost DACs. This is because power consumption due to quantization increases exponentially as resolution increases. Various non-linear precoding methods with phase-shift-keying (PSK) constellations have been studied actively in the system, which can achieve good performances with low-complexities \cite{park2019construction,li2018massive,castaneda20171,landau2017branch}. However, the above methods cannot be straightforwardly applied to more practical quadrature-amplitude-modulation (QAM) constellations since in QAM constellations, the decision regions are bounded. Recently, various precoding methods for QAM constellations have been investigated in  \cite{amor201716,jacobsson2016nonlinear,sohrabi2018one,jacobsson2017quantized,jedda2018quantized}. Leveraging the fact that 16-QAM symbols can be obtained as the superposition of two QPSK symbols, the authors in \cite{amor201716} formulated an optimization problem using gradient projection to obtain 1-bit transmit vectors, and stored them in a look-up-table per coherent channel. In \cite{jacobsson2016nonlinear}, non-linear 1-bit precoding methods for massive MIMO with QAM constellations have been proposed, which are enabled by semi-definite relaxation and $\ell_\infty$-norm relaxation. However, these methods do not provide an elegant complexity-performance trade-off. Thus, it is necessary to investigate a precoding method with an attractive performance and low-complexity under QAM constellations, which is the major subject of this paper.
\textcolor{blue}{Some precoding methods with 1-bit DACs follow the minimum mean square error (MMSE) design criterion. In \cite{castaneda20171}, C1PO method is proposed from using bi-convex relaxation. Due to complexity of matrix inversion at C1PO, C2PO has been proposed as a low-complexity algorithm variant.
The paper shows attractive error-rate performance with efficient complexity at low-order modulation. In \cite{chen2019mmse}, The MMSE-based one-bit precoding, MMSE-ERP is developed with enhanced receive processing capability at the mobile stations. the algorithm is based on a combination of the alternating minimization method using a projected gradient method, and equilibrium constraint. the performance of MMSE-ERP is significant. Also, \cite{wang2018finite} provides IDE algorithm that exploits an alternating direction method of multipliers (ADMM) framework. Furthermore, complexity efficient algorithm called as a IDE2. Both of IDE and IDE2 achieve excellent error-rate performance. 
constructive interference (CI) design criterion is similar to our one. \cite{li2020interference, li20201bit,li2018massive} introduce many symbol-level precoding methods that have CI design critrion. Symbol scaling is the efficient algorithm that achieve good performance with PSK. In \cite{li20201bit,li2020interference}, the optimization problems are defined with both of equality constraints and inequality constraints. Based on the problems, A partial branch and bound(P-BB) and an ordered partial sequential update(OPSU) achieve near-optimal performance and significant performance, respectively.
In \cite{jedda2018quantized}, maximum safety margin (MSM) design criterion exploiting the constructive interference. Also, MSM algorithm and analysis of the algorithm are provided for constant envelope precoding with PSK and QAM.}

In this paper, we present a novel direction to construct a 1-bit transmit signal vector for a downlink MU-MISO system with 1-bit DACs. Toward this, our contributions are summarized as follows:
\begin{itemize}

\item We first derive the so-called {\em feasibility condition} which ensures that each user's noiseless observation is placed in a desired decision region. That is, if a transmit signal is constructed to satisfy the feasibility condition, each user can recover a desired signal successfully in a higher signal-to-noise ratio (SNR). 

\item Incorporating the robustness to an additive noise into the feasibility condition, we show that our problem to construct a 1-bit transmit signal vector can be formulated as a mixed integer linear programming (MILP). This problem can be optimally solved via a linear programming (LP) relaxation and branch-and-bound. 

\item Furthermore, we propose a low-complexity algorithm to solve the MILP, by introducing a novel greedy algorithm, which can almost achieve the optimal performance with much lower computational complexity.


\item Via simulation results, we demonstrate that the proposed method can outperform the state-of-the-art methods. Furthermore, the complexity comparisons of the proposed and existing methods demonstrate the potential of the proposed direction and algorithm.

 
\end{itemize}

This paper is organized as follows. In Section~\ref{sec:pre}, we provide useful notations and definitions, and describe a system model. In Section \ref{sec:structure}, we propose an efficient method to construct a transmit signal vector for downlink MU-MISO systems with 1-bit DACs. Moreover, the low complexity methods are proposed in Section \ref{sec:low complexity methods}. Section \ref{sec:simulation} provides simulation results. Conclusions are provided in  \ref{sec:conclusion}. 
\section{Preliminaries}\label{sec:pre}

In this section, we provide useful notations used throughout the paper, and then describe the system model.

\subsection{Notation}
The lowercase and uppercase bold letters represent column vectors and matrices, respectively. The symbol $(\cdot)^{\transp}$ denotes the transpose of a vector or a matrix. For any vector $\bf x$, $x_i$ represents the $i$-th component of $\bf x$. Let $\left[a:b\right]\eqdef\{a,a+1,\ldots,b\}$ for any integer $a$ and $b$ with $a<b$. The notation of $\text{card}(\mathcal{U})$ denotes the number of elements of a finite set  $\mathcal{U}$. A rank of a matrix \Am \ is represented as rank(\Am). 
$\Re(\av)$ and $\Im(\av)$ represent the real and complex parts of a complex vector $\av \in \CC$, respectively.
 For any $x\in\mathbb{C}$, we let
 \begin{equation}\label{eq:1}
     g(x)=[\Re(x), \Im(x)]^{\transp},                          
 \end{equation} 
 and the inverse mapping of $g$ is denoted as $g^{-1}$. Also, $g$ and $g^{-1}$ are the component-wise operations, i.e., $g([x_1,x_2]^{\transp})=[\Re(x_1),\Im(x_1),\Re(x_2),\Im(x_2)]^{\transp}$. For a complex-value $x$, its real-valued matrix expansion $\phi(x)$ is defined as
 \begin{equation}\label{eq:2}
     \phi(x)=\left[{\begin{array}{cc}
     \Re(x)&  -\Im(x)\\
     \Im(x)&  \Re(x)
     \end{array}}\right].
 \end{equation}
 As an extension into a vector, the operation of $\phi$ is applied in an element-wise manner as
 \begin{equation}
     \phi([x_1,x_2]^{\transp})=[\phi(x_1)^{\transp},\phi(x_2)^{\transp}]^{\transp}.
 \end{equation}
 $\bar{\bf{1}}_n$ denotes the length-$n$ all-one vector, and $\otimes$ indicates Kronecker product operator.


\subsection{System Model}
We consider a downlink MU-MISO system where BS equipped with $N_t \gg K$ transmits antennas serves $K$ users with a single antenna. As a natural extension of our earlier work in \cite{park2019construction} which focuses on PSK constellations only, this paper considers $4^n$-QAM with $n\geq 2$. Let $\Cc$ denote the set of constellation points of $4^n$-QAM. Also, letting  $\xv=[x_1,\ldots,x_{N_t}]^{\transp}$ be a transmit vector at the BS,  the received signal vector 
$\xv\in\CC^{K}$ at the $K$ users is given as
 \begin{equation}\label{eq:3}
     \yv = \sqrt{\rho}\Hm\xv+\zv,
 \end{equation}
 where $\Hm\in\mathbb{C}^{K\times N_t}$ denotes the frequency-flat Rayleigh fading channel, each of which component follows a complex Gaussian distribution with zero mean and unit variance, and  $\zv\in\mathbb{C}^{K\times 1}$ denotes the additive Gaussian noise vector whose each element are distributed as complex Gaussian random variables with zero mean and unit variance, i.e., $z_i \sim \mathcal{CN}(0,\sigma^2=1)$. The SNR is defined as $\SNR=\rho/\sigma^2$, where $\rho$ denotes the per-antenna power constraint. Throughout the paper, it is assumed that the channel matrix $\Hm$ is perfectly known at the BS.
 
 Given a message vector $\sv\in\mathcal{C}^K$, BS needs to construct a transmit vector $\xv$ such that each user $k$ can recover the desired message $s_k$ successfully. Toward this, our goal is to construct such a precoding function $\Pc$:
 \begin{equation}\label{eq:4}
     \xv=\mathcal{P}(\Hm,\sv),
 \end{equation} which produces a transmit vector $\xv$ from the channel matrix $\Hm$ and the message vector $\sv$. Focusing on the impact of 1-bit DACs on the downlink precoding, we assume that BS is equipped with 1-bit DACs while all $K$ users are equipped with infinite-resolution ADCs. 
To isolate the performance impact of 1-bit DACs, 
each component $x_i$ of the transmit vector $\xv$ is restricted as 
 \begin{equation}\label{eq:5}
     \Re(x_i)\  \rm{and}\ \Im(\mathit{x_i})\in \{-1,1\}.
 \end{equation} Since this restriction causes a severe non-linearity, conventional precoding methods, developed by exploiting the linearity, cannot ensure an attractive performance. The objective of this paper is to construct a precoding function $\mathcal{P}(\Hm,\sv)$ with a manageable complexity and suitable for the considered non-linear MISO channels.



\section{The Proposed Transmit-Signal Vectors}\label{sec:structure}

\begin{figure}[!t]
    \begin{center}
    \includegraphics[width=0.35\textwidth]{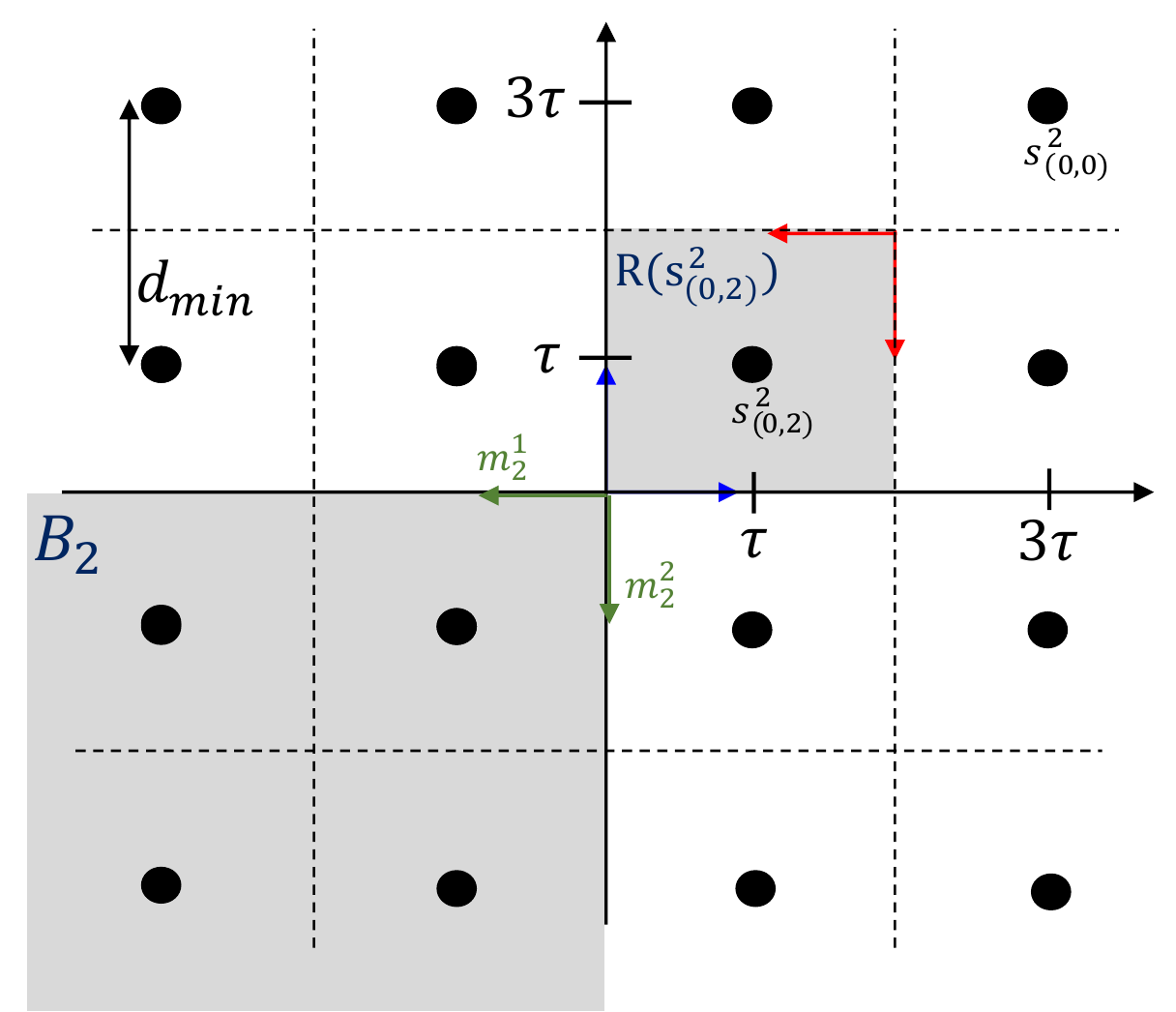}
    \end{center}
    \caption{Description of the decision regions for $4^2$-QAM with adaptive $\tau$.}
    \label{fig:2}
\end{figure}

We formulate an optimization problem to construct a transmit-vector $\xv$ under $4^n$-QAM. Especially, this problem can be represented as a manageable MILP. We remark that our earlier work on PSK  \cite{park2019construction} cannot be employed as the decision regions of $4^n$-QAM because a QAM constellation is bounded (see Fig.~\ref{fig:2}). For the ease of exploration, an equivalent real-valued expression is used as follows:
\begin{equation}\label{eq:9}
         \tilde{\yv} = \sqrt{\rho}\tilde{\Hm}\tilde{\xv}+\tilde{\zv},
\end{equation}
where $\tilde{\xv}=g(\xv)$, $\tilde{\xv}=g(\xv)$, $\tilde{\zv}=g(\zv)$, and $\tilde{\Hm}=\phi(\Hm)\in\mathbb{R}^{2K\times 2N_t}$ denotes the real-value expansion matrix of $\Hm$.

Before explaining the main result, we provide the useful definitions which are used throughout the paper.

\vspace{0.1cm}
\begin{definition}\label{def1}{\em (Decision region)} For any constellation point $s\in\mathcal{C}$, the decision region of $s$ is defined as
\begin{equation}\label{eq:6}
    \mathcal{R}(s) \triangleq \left\{y\in\mathbb{C}:|y-s| \le \min_{c\in\mathcal{C}:c\ne s}|y-c|\right\}.
\end{equation} This region means that a received signal $y \in\mathcal{R}(s)$ is detected as $s$. In addition the real-valued decision region is given as
\begin{equation}
    \tilde{\mathcal{R}}(s)=g\left(\mathcal{R}(s)\right).
\end{equation}\hfill$\blacksquare$
\end{definition} 

\vspace{0.1cm}
\begin{definition}\label{def2}{\em (Base region)} A base region $\tilde{\mathcal{B}}_i\subseteq \mathbb{R}^2, \forall i\in\left[0:3\right]$, is defined as
\begin{equation}\label{eq:7}
    \tilde{\mathcal{B}}_{i} \triangleq \{\alpha_{i}^1\mv_{i}^1+\alpha_{i}^2\mv_{i}^2: \alpha_{i}^1,\alpha_{i}^2>0\},
\end{equation}
where $\mv_i^\ell$ represents a basis vector with
\begin{equation}\label{eq:8}
    \mv_{i}^{\ell} = \begin{cases} g\left(\sqrt{2}\cos(\frac{\pi}{4}(1+2i))\right) & \mbox{if }\ell=1 \\ g\left(j\sqrt{2}\sin(\frac{\pi}{4}(1+2i))\right) & \mbox{if }\ell=2.  \end{cases}
\end{equation}\hfill$\blacksquare$
\end{definition} 
\vspace{0.1cm}
\begin{definition}\label{def3}{\em (Partial matrix)} A partial matrix $\Am_{\mathcal{U}}\in\mathbb{R}^{m\times \text{card}(\mathcal{U})}$ is defined as 
\begin{equation}
    \Am_{\mathcal{U}}\triangleq[(\Am^{\transp})_{u_1},(\Am^{\transp})_{u_2},\ldots,(\Am^{\transp})_{u_{\text{card}(\mathcal{U})}}]^{\transp},
\end{equation}
where $\Am\in\mathbb{R}^{m\times n}$, $\mathcal{U}\triangleq \{u_1,u_2,\ldots,u_{\text{card}(\mathcal{U})}\} \subseteq[1:n]$ and $(\Am^{\transp})_k$ denotes the k-th row of $\Am^{\transp}$.
\end{definition}
\hfill$\blacksquare$
\vspace{0.1cm}
\begin{definition}\label{def4}{\em (Partial vector)} A partial vector $\xv_{\mathcal{U}}\in\mathbb{R}^{\text{card}(\mathcal{U})\times 1}$ is defined as 
\begin{equation}
   \xv_{\mathcal{U}}\triangleq[x_{u_1},x_{u_2},\ldots,x_{u_\text{card}(\mathcal{U})}]^{\transp},
\end{equation}
where $\xv\in\mathbb{R}^{n\times 1}$, $\mathcal{U}\triangleq \{u_1,u_2,\ldots,u_{\text{card}(\mathcal{U})}\} \subseteq[1:n]$.
\end{definition} 
\vspace{0.1cm}

In the sequel, the decision region in Definition~\ref{def1} will be represented by the intersections of the $n$ base regions in Definition~\ref{def2} with proper shift values. This representation makes it easier to formulate an optimization problem. First of all, we need to decide the size of bounded decision regions, i.e., the parameter $\tau$ in Fig.~\ref{fig:2} should be determined. Note that $2\tau=d_{\rm min}$ denotes the minimum Euclidean distance of the given constellation points. In PSK, $\tau$ is always infinite regardless of a channel matrix, whereas in $4^n$-QAM, it should be well-optimized. 
Specifically, $\tau$ should be chosen as large as possible to ensure a reliable performance, provided that a noiseless received signal belongs to the corresponding decision regions at all the $K$ users. 




From now on, we will explain how to construct a transmit-signal vector $\xv$ for a decision-size $\tau$. Throughout the paper, it is assumed that all $K$ users have the identical decision size $\tau$ for the practicability of an optimization. Given $4^n$-QAM, each symbol is indexed by a length-$n$ quaternary vector $(i_1,...,i_n)$ with $i_j \in [0:3]$, i.e., 
\begin{equation}
\mathcal{C}=\left\{s_{(0,\ldots,0)}^{n},s_{(0,\ldots,1)}^{n},\ldots,s_{(3,\ldots,3)}^{n}\right\}.
\end{equation} Each constellation point can be represented as a linear combination of the $n$ basis symbols $c_i$'s such as 
 \begin{equation}\label{eq:13}
     s_{(i_1,\ldots, i_n)}^{n} \triangleq \tau \sum_{l=1}^n 2^{n-l} c_{i_l} = \tau {s'}_{(i_1,\ldots, i_n)}^{n}.
 \end{equation} Here, a normalized constellation point and the basis symbols are defined as
\begin{equation}\label{eq:12}
    {s'}_{(i_1,\ldots, i_n)}^{n} \triangleq \sum_{l=1}^n 2^{n-l}  c_{i_l}, 
\end{equation}
where $c_i$'s, four fundamental constellation points are given as
\begin{equation}
    c_i \triangleq \sqrt{2}\left\{{\cos\left({\frac{\pi}{4}} (1+2i)\right)+j\sin\left({\frac{\pi}{4}} (1+2i)\right)}\right\},
\end{equation} 
for $i\in[0:3]$. For the ease of expression, we represent the constellation $\mathcal{C}$ and the corresponding decision regions  $\mathcal{R}( s^{n}_{(i_1,\ldots, i_n)})$ in the corresponding real-valued forms:
\begin{equation}\label{eq:14}
    \tilde{\mathcal{C}} = \left\{g(s_{(0,\ldots,0)}^{n}),g(s_{(0,\ldots,1)}^{n}),\ldots, g(s_{(3,\ldots, 3)}^{n})\right\},
\end{equation} and
\begin{equation}\label{eq:15}
    \tilde{\mathcal{R}}\left( s^{n}_{(i_1, \ldots, i_n)}\right) = g\left(\mathcal{R}\left( s^{n}_{(i_1, \ldots ,i_n)}\right)\right).
\end{equation} 
A transmit vector $\xv$ should ensure that a noiseless received signal at the $k$-th user (i.e., $r_k=\hv_k^T\xv$) should be placed in the corresponding decision regions for all users $k\in[1:K]$. This necessity condition implies that  $\xv$ should satisfy the following condition:
\begin{align}\label{eq:region_constraint}
     g(r_k) &\in \tilde{\mathcal{R}}\left( s^{n}_{(\mu_{k,1},\ldots, \mu_{k,n})}\right), k \in [1:K],
\end{align} for $k \in [1:K]$, where $r_k=\hv_k^T\xv$ denotes a noiseless received signal (i.e., $y_k =\hv_k^T\xv +z_k$).

\vspace{0.2cm}
\noindent{\bf  Feasibility condition:}  The condition in (\ref{eq:region_constraint}) can be rewritten in a way that the optimization problem can be interpreted as an LP problem. The decision region in (\ref{eq:region_constraint}) can be expressed as the intersections of the $n$ shifted base regions in Definition~\ref{def2}:
\begin{align}\label{eq:18}
  \tilde{\mathcal{R}}\left( s^{n}_{(i_1, \ldots, i_n)}\right)\triangleq \tilde{\mathcal{B}}_{i_1}\bigcap_{l=2}^n \left\{\tilde{\mathcal{B}}_{i_l}+2^{n-(l-1)}g\left(s_{(i_1,\ldots, i_{l-1})}^{l-1}\right)\right\},
\end{align}where the shifted base region with a bias $c$ is defined as
\begin{equation}\label{eq:19}
    \tilde{\mathcal{B}}_{i}+c \triangleq \{\alpha_{i}^1\mv_{i}^1+\alpha_{i}^2\mv_{i}^2+c: \alpha_{i}^{1},\alpha_{i}^{2}>0\}.
\end{equation} Then, the condition in (\ref{eq:region_constraint}) holds if  $g(r_k)$ can be represented by the following $n$ linear equations with some positive coefficients, i.e.,
\begin{align}\label{eq:20}
    g(r_k) &= \alpha_{k,1}^1\mv_{\mu_{k,1}}^1+\alpha_{k,1}^2\mv_{\mu_{k,1}}^2+2^{n}g(0) \\
            & = \alpha_{k,2}^1\mv_{\mu_{k,2}}^1+\alpha_{k,2}^2\mv_{\mu_{k,2}}^2+2^{n-1}g(s_{(\mu_{k,1})}^{1}) \nonumber  \\
            &\;\; \vdots  \nonumber \\
            & = \alpha_{k,n}^1\mv_{\mu_{k,n}}^1+\alpha_{k,n}^2\mv_{\mu_{k,n}}^2+2^1g(s_{(\mu_{k,1},\ldots ,\mu_{k,n-1})}^{n-1}), \nonumber
\end{align} for some $\alpha_{k,1}^1,\alpha_{k,1}^2, \ldots,\alpha_{k,n}^1,\alpha_{k,n}^2 \ge 0$.
The condition in (\ref{eq:20}) is called a {\em feasibility} condition as it can guarantee that $r_k\in \mathcal{R}\left( s^{n}_{(\mu_{k,1},\ldots,\mu_{k,n})}\right)$ for $k\in [1:K]$. In other words, if this condition is satisfied, all $K$ users can reliably detect the desired messages in higher SNRs.

\begin{example}Assuming $4^2$-QAM, we explain how to obtain the feasibility condition in (\ref{eq:18}).
Consider the decision region $\mathcal{R}(s^{2}_{(0,2)})$. From Fig.~\ref{fig:2}, the decision region is represented by the intersection of the two base regions $\mathcal{B}_0$ (i.e., the infinite region with blue basis in Fig.~\ref{fig:2}) and {$\mathcal{B}_2+s_{(0)}^{1}$} (i.e., the infinite region with red basis in Fig.~\ref{fig:2}). Thus,  the decision region (i.e., the gray region in Fig.~\ref{fig:2}) is represented as
\begin{equation}\label{eq:21}
    \mathcal{R}\left(s^{2}_{(0,2)}\right) \triangleq \left\{\mathcal{B}_{0}+2^2g(0)\right\}\cap \left\{\mathcal{B}_{2}+2^1s_{(0)}^{1}\right\}.
\end{equation} Also, from Definition 2, the above condition can be represented by the following two linear equations:
\begin{align}\label{eq:23}
    g(\rv_k) =& \alpha_{k,1}^1\mv_{0}^1+\alpha_{k,1}^2\mv_{0}^2+2^2g(0),\nonumber \\ 
             = &\alpha_{k,2}^1\mv_{2}^1+\alpha_{k,2}^2\mv_{2}^2+2^1 g\left(s_{(0)}^{1}\right),
\end{align} for some positive coefficients $\alpha_{k,1}^1,\alpha_{k,1}^2,\alpha_{k,2}^1,\alpha_{k,2}^2>0$. This is equivalent to the condition in (\ref{eq:20}). In the same way, we can verify the feasibility condition in (\ref{eq:18}).   \hfill$\blacksquare$
\end{example}
\vspace{0.1cm}

We are now ready to derive MILP problem which can generate a good transmit vector $\xv$ under 1-bit DAC constraints. We first represent the feasibility condition in a matrix form. Define the $n$ copies of the channel vector $\hv_k$ as
\begin{equation}
    \Hm^k\eqdef \bar{\bf{1}}_n \otimes \hv_k=[\underbrace{\hv_k^\transp,\ldots, \hv_k^\transp}_{n}]^\transp, \label{eq:26} 
\end{equation} where $\hv_k$ denotes the $k$-th row of $\Hm$. Also, the corresponding real-valued expression is denoted as 
\begin{equation}
    \tilde{\Hm}^k=\phi(\Hm^k).
\end{equation} Accordingly, the $n$-extended received vector at $k$-th user is written as
\begin{align}\label{eq:27}
    \rv^k&\triangleq g(\Hm^k\xv) \nonumber\\
    &=\tilde{\Hm}^k\tilde{\xv} =\bar{\bf{1}}_n \otimes g(r_k).
\end{align} We next express the right-hand side of \eqref{eq:20}, i.e., linear constraints, in a matrix form.  From Definition 2, we let:
\begin{align}
    \Mm_i \triangleq [\mv_{i}^1 \ \mv_{i}^2]&= \begin{bmatrix} \Re(c_{i}) & 0 \\ 0 & \Im(c_{i})   \end{bmatrix} \nonumber\\
    &= \begin{bmatrix} \sqrt{2}\cos\left({\frac{\pi}{4}} (1+2i)\right) & 0 \\ 0 & \sqrt{2}\sin\left({\frac{\pi}{4}} (1+2i)\right)  \end{bmatrix}. \label{eq:25}     
\end{align} We notice that $\Mm_i$ is a symmetric and orthogonal matrix as
\begin{align}\label{eq:sym_or}
    \Mm_{i}\Mm_{i}
    = \begin{bmatrix} \cos\left({\frac{\pi}{2}} (1+2i)\right)+1 & 0 \\ 0 & -\cos\left({\frac{\pi}{2}} (1+2i)\right)+1  \end{bmatrix}=\Id. 
\end{align}
Since the decision region of a constellation point $4^n$-QAM is formed as the conjunction of $n$ shifted base regions, we need to establish a tightly packed format that can cope with both base regions and shifts (biases, equivalently). The former is addressed by the basis matrix $\Mm^{\mu_k}$ and coefficient vector $\alphav^k$, which are respectively written as 
\begin{align}
    \Mm^{\mu_k} &\triangleq \mbox{diag}(\Mm_{\mu_{k,1}},\ldots,\Mm_{\mu_{k,n}}) \label{eq:28}\\
\alphav^k &\triangleq[\alpha_{k,1}^1,\alpha_{k,1}^2,\ldots,\alpha_{k,n}^1,\alpha_{k,n}^2]^\transp.
\end{align}Lastly, the whole series of the biases are formed as the normalized bias vector $\bv$ with $\tau$, which is defined from (\ref{eq:13}) as
 \begin{align}\label{eq:29}
    \bv^{\mu_k} \triangleq g\left([2^n\cdot0, 2^{n-1}\cdot {s'}_{(\mu_{k,1})}^{1}, \ldots ,2^1\cdot {s'}_{(\mu_{k,1},\ldots,\mu_{k,n-1})}^{n-1}]^\transp\right)\nonumber\\
    =\frac{1}{\tau}g\left([2^n\cdot0, 2^{n-1}\cdot s_{(\mu_{k,1})}^{1}, \ldots ,2^1\cdot s_{(\mu_{k,1},\ldots,\mu_{k,n-1})}^{n-1}]^\transp\right).
 \end{align}
 
From \eqref{eq:28}-\eqref{eq:29}, the matrix form of $k$-th user's feasibility conditions (\ref{eq:20}) is given as
\begin{equation}\label{eq:31}
   \rv^k=\Mm^{\mu_k}\boldsymbol{\alpha}^k+\tau\bv^{\mu_k}.
\end{equation}
Leveraging the expression designed for each user, we construct the cascaded matrix form of feasibility conditions on all $K$ users as
\begin{equation}\label{eq:32}
    \Bar{\rv}=\Bar{\Hm}\tilde{\xv}=\Bar{\Mm}\bar{\boldsymbol{\alpha}}+\tau\Bar{\bv},
\end{equation}
where 
\begin{align*}
        \Bar{\Mm} &\triangleq \mbox{diag}(\Mm^{\mu_1},\ldots,\Mm^{\mu_K}),\; \Bar{\Hm} \triangleq [(\tilde{\Hm}^{1})^\transp,\ldots,(\tilde{\Hm}^{K})^\transp]^\transp \\
 \Bar{\rv} &\triangleq [(\rv^{1})^\transp,\ldots,(\rv^{K})^\transp]^\transp,\; \Bar{\bv} \triangleq [(\bv^{\mu_1})^\transp,\ldots,(\bv^{\mu_K})^\transp]^\transp \\
        \bar{\boldsymbol{\alpha}}&\triangleq [(\boldsymbol{\alpha}^1)^\transp,\ldots,(\boldsymbol{\alpha}^K)^\transp]^\transp.
\end{align*} 
Thus, the feasibility condition in (\ref{eq:32}) is rewritten as
\begin{equation}\label{eq:37}
    \bar{\boldsymbol{\alpha}}=\underbrace{\bar{\Mm}\bar{\Hm}}_{\triangleq{\boldsymbol{\Lambda}}}\tilde{\xv}-\tau\underbrace{\bar{\Mm}\bar{\bv}}_{\triangleq{\boldsymbol{\Lambda}_b}},
\end{equation} 
using the fact that $\bar{\Mm}^{-1}=\bar{\Mm}$ from (\ref{eq:sym_or}). We remark that $\boldsymbol{\Lambda}\in\mathbb{R}^{2nK\times2N_t}$ and $\boldsymbol{\Lambda}_b\in\mathbb{R}^{2nK\times1}$ are fully determined by the channel matrix $\bar{\Hm}$ and users' messages $\{\mu_k: k\in[1:K]\}$. 

\noindent{\bf Robustness:} A feasible transmit vector can provide an attractive performance in higher SNR regimes. However, the robustness to an additive Gaussian noise cannot be guaranteed. To enhance the robustness, one reasonable way is to make a noiseless received signal to be placed in the center of the decision region. Namely, we aim at moving away the noiseless signal from the boundaries of the decision areas. By taking this goal into account, we formulate the following optimization problem:
\begin{align}\label{eq:39}
 &\mathcal{P}_1:&& \max_{\tilde{\xv},\tau} \min\{\alpha_{k,j}^i: i=1,2,\ j\in[1:n],\ k\in[1:K]\} \\
  & \text{s.t.} &&\bar{\boldsymbol{\alpha}}=\boldsymbol{\Lambda}\tilde{\xv}-\tau\boldsymbol{\Lambda}_{b},\nonumber\\
  &&& \alpha_{k,j}^1,\alpha_{k,j}^2 > 0,\ j\in[1:n],\ k\in[1:K],\nonumber\\
   &&& \tilde{\xv}\in\{-1,1\}^{2N_t} \nonumber.
\end{align}
\begin{figure}[!t]
    \centering
    \includegraphics[width=0.5\textwidth]{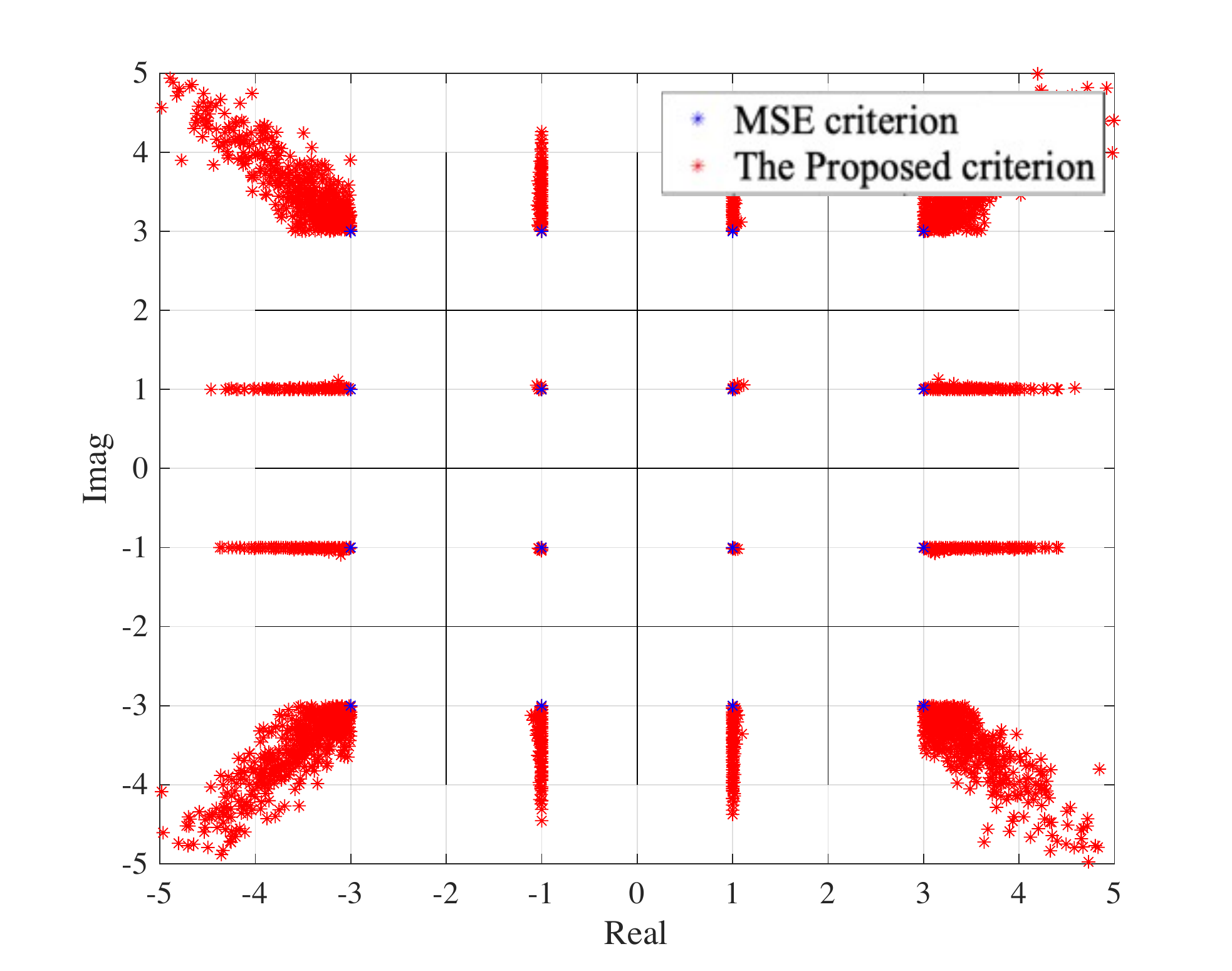}
    \caption{The normalized noiseless received signals of $4^2$-QAM.}
    \label{fig:4}
\end{figure} 


From now on, we explain how to solve the problem $\mathcal{P}_1$ efficiently. According to dealing with the decision parameter $\tau$, we consider the two different scenarios: i) a fixed $\tau$ irrespective of a channel matrix $\Hm$; ii) a channel-dependent $\tau$. Clearly, the second scenario requires a more overhead since in this case, BS needs to transmit the $\tau$ to the $K$ users more frequently. For the first scenario, we employ  the asymptotic result provided in \cite{sohrabi2018one}, where it is fully determined as a function of $N_t$, $n$ and $K$:
\begin{equation}\label{eq:10}
     \tau \eqdef {\frac{\sqrt{{2/ \pi}}}{6}} \sqrt{\frac{2\rho {N_t}^2}{\tilde{f}(K,n)}},
 \end{equation}
 where
 \begin{equation}\label{eq:11}
     \tilde{f}(K,n) = K\frac{2^n+1}{3(2^n-1)}+2\sqrt{K\frac{(2^n+1)(2^{2n}-4)}{ 22.5(2^n-1)^3}}.
 \end{equation} Leveraging the fixed $\tau$ in (\ref{eq:10}), $\mathcal{P}_1$ can be formulated as MILP: 
\begin{align}\label{eq:43}
 &\mathcal{P}_2:&& \argmax_{\tilde{\xv},t}\ \  t \nonumber\\
  & \text{s.t.} &&\boldsymbol{\Lambda}_i\tilde{\xv}-\tau \boldsymbol{\Lambda}_{b,i} \ge t,\ i\in[1:2nK], \nonumber\\ 
  &&& t>0, \nonumber\\
   &&& \tilde{\xv}\in\{-1,1\}^{2N_t},
\end{align} where $\boldsymbol{\Lambda}_i$ and $\boldsymbol{\Lambda}_{b,i}$ denote the $i$-th row of  $\boldsymbol{\Lambda}$ and $\boldsymbol{\Lambda}_b$, respectively. For the second scenario, $\tau$ is set by $\tau \eqdef t$. In general, this value can be changed according to a channel matrix $\Hm$. This choice is motivated by the fact that $t$ is the lower bound of the coefficients $\bar{\alphav}$, and due to the normalized $\bar {\Mm}$ and $\bar{\bv}$, the coefficients directly signify how far away it is from a detection boundary. Accordingly, the optimization problem to find the decision parameter $\tau$ and transmit vector $\xv$ simultaneously is defined as
\begin{align}\label{MILP4}
 &\mathcal{P}_3:&& \argmax_{\tilde{\xv},t}\ \  t \nonumber\\
  & \text{s.t.} &&\frac{1}{1+\boldsymbol{\Lambda}_{b,i}}\boldsymbol{\Lambda}_i\tilde{\xv} \ge t,\ i\in[1:2nK], \nonumber\\ 
  &&& t>0, \nonumber\\
   &&& \tilde{\xv}\in\{-1,1\}^{2N_t}.
\end{align} 
{\BLUE The proposed MILP problems in $\mathcal{P}_2$ and $\mathcal{P}_3$ can be solved via the well-known branch-and-bound (B\&B) method like in \cite{landau2017branch}, when implemented correctly, should identify the optimal precoding vector in its criterion. Furthermore, the partial B\&B like in \cite{li2020interference} is, as the title suggest, a nearly optimal method.
} However, its computational complexity is quite expensive for a realistic implementation \cite{landau2017branch}. In the following section, we will present a low-complexity method to solve the optimization problems in $\mathcal{P}_2$ and $\mathcal{P}_3$ efficiently.



\begin{remark} Fig.~\ref{fig:4} verifies the proposed approach, where $10^4$ normalized noiseless signals, i.e.,  $\Hm\xv$, are plotted with $N_t=8$, $K=2$, and $n=2$. The blue points depict the noiseless received signals when ZF precoding in \cite{peel2005vector} is used with the assumption of infinite resolution. In contrast, the red points show the noiseless received signals when the proposed 1-bit transmit vectors obtained from the solutions of $\mathcal{P}_2$ are used. Fig.~\ref{fig:4} clearly shows that the red points can provide more robustness than the blue points even with the low-resolution data converters. \hfill$\blacksquare$
\end{remark}

\section{Low-Complexity Precoding Methods}\label{sec:low complexity methods}
In this section, we present an efficient algorithms to solve MILP problems in $\mathcal{P}_2$ and $\mathcal{P}_3$. 
We first  solve the LP problem by relaxing the integer constraint in $\mathcal{P}_3$ as the bounded interval:
\begin{align}\label{eq:46}
 &\mathcal{P}_4:&& \argmax_{\tilde{\xv},t}\ \  t \nonumber\\
  & \text{s.t.} &&\frac{1}{1+\boldsymbol{\Lambda}_{b,i}}\boldsymbol{\Lambda}_i\tilde{\xv} \ge t,\ i\in[1:2nK], \nonumber\\ 
  &&& t>0, \nonumber\\
  &&& -1\le\tilde{x}_j\le 1,\ j\in[1:2N_t].
\end{align}
Similarly, $\mathcal{P}_2$ can reformed as the following LP problem with relaxed linear constraint:
\begin{align}\label{LP5}
 &\mathcal{P}_5:&& \argmax_{\tilde{\xv},t}\ \  t \nonumber\\
  & \text{s.t.} &&\boldsymbol{\Lambda}_i\tilde{\xv}-\tau \boldsymbol{\Lambda}_{b,i} \ge t,\ i\in[1:2nK], \nonumber\\ 
  &&& t>0, \nonumber\\
  &&& -1\le\tilde{x}_j\le 1,\ j\in[1:2N_t],
\end{align} where $\tau$ is given in (\ref{eq:10}).

The above problems can be efficiently solved via simplex method \cite{luenberger1984linear}, and the corresponding relaxed LP solutions are denoted as $\tilde{\xv}_{\rm LP}$. We then refine the solution of $\mathcal{P}_4$ or $\mathcal{P}_5$ via a greedy algorithm (see Algorithm 1), so that it satisfies the one-bit constraints. Starting from the solutions of $\mathcal{P}_4$ or $\mathcal{P}_5$, i.e., $\tilde{\xv}_{\rm LP}$, the main steps behind the second stage are 1) choosing an antenna index $i$; 2) testing the possible values of the antennas, that is $\tilde{x}_i \in \{-1,1\}$; 3) calculating the set of scaling coefficients when artificially changing $\tilde{x}_i$; and 4) finally setting $\tilde{x}_i=j$ where the substitution of $j\in\{-1,1\}$ for $\tilde{x}_i=j$ insists the maximization of the minimum element in the coefficients.
\begin{algorithm}[t]\label{al:1}
\caption{Greedy Algorithm}
\textbf{Input:}  $\tilde{\xv}_{\rm LP}\in\mathbb{R}^{2N_t\times1}$, $\boldsymbol{\Lambda}\in\mathbb{R}^{2nK\times2N_t}$, $\boldsymbol{\Lambda}_b\in\mathbb{R}^{2nK\times1}$ and $\tau\in\mathbb{R}^{+}$.

\textbf{Initialization:} $\tilde{\xv}=\tilde{\xv}_{\rm LP}$ (obtained by either $\mathcal{P}_4$ or $\mathcal{P}_5$).
\begin{algorithmic}
\For{$i=1:2N_t$}
\For{$j\in \{-1,1\}$}
\State $\tilde{x}_i=j$ and $\bar{\boldsymbol{\alpha}}^{(j)}=\boldsymbol{\Lambda}\tilde{\xv}-\tau \boldsymbol{\Lambda}_b$
\EndFor
\State  Update $\tilde{x}_i\leftarrow \argmax_{j\in\{-1,1\}}\{\min(\bar{\boldsymbol{\alpha}}^{(j)})\}$
\EndFor
\State \textbf{Output:} $\tilde{\xv}\in\mathcal{R}^{2N_t\times1}$
\end{algorithmic}
\end{algorithm}
%

\begin{algorithm}[t]\label{al:2}
\caption{\textcolor{blue}{Partial Greedy Algorithm}}
\textbf{Input:}  $\tilde{\xv}_{\rm LP}\in\mathbb{R}^{2N_t\times1}$, $\boldsymbol{\Lambda}\in\mathbb{R}^{2nK\times2N_t}$, $\boldsymbol{\Lambda}_b\in\mathbb{R}^{2nK\times1}$, $\tau\in\mathbb{R}^{+}$, $\mathcal{O}\subseteq[1:2N_t]$, $\mathcal{Q}\subseteq[1:2N_t]$.

\textbf{Initialization:} $\tilde{\xv}=\tilde{\xv}_{\rm LP}$ (obtained by either $\mathcal{P}_4$ or $\mathcal{P}_5$),\\ $\boldsymbol{\Lambda}_k=\tau\boldsymbol{\Lambda}_b-\boldsymbol{\Lambda}_{\mathcal{O}}\tilde{\xv}_{\mathcal{O}}$.
\begin{algorithmic}
\For{$i=1:\text{card}(\mathcal{Q})$}
 \For{$j\in \{-1,1\}$}
\State $\tilde{x}_{\mathcal{Q}_i}=j$ and $\bar{\boldsymbol{\alpha}}^{(j)}=\boldsymbol{\Lambda}_{\mathcal{Q}}\tilde{\xv}_{\mathcal{Q}}- \boldsymbol{\Lambda}_k$
\EndFor
\State  Update $\tilde{x}_{\mathcal{Q}_i}\leftarrow \argmax_{j\in\{-1,1\}}\{\min(\bar{\boldsymbol{\alpha}}^{(j)})\}$
\EndFor
\State \textbf{Output:} $\tilde{\xv}\in\mathcal{R}^{2N_t\times1}$
\end{algorithmic}
\end{algorithm}

\subsection{Greedy algorithms}
$\tilde{\xv}_{\rm LP}$, which is the solution of $\mathcal{P}_4$ or $\mathcal{P}_5$, is obtained via simplex method \cite{luenberger1984linear} instead of the interior point method \cite{den2012interior}. The simplex method concentrates on finding out an optimal solution that is an extreme point of constraint set of $\mathcal{P}_4$ or $\mathcal{P}_5$. Via numerical tests, we have confirmed that the extreme points are basic feasible solutions whose most entries are already satisfying 1-bit constraint. A theoretical proof is non-trivial and left for an interesting future work.



From $\tilde{\xv}_{\rm LP}$, the solution of LP, set $\mathcal{O}$ and $\mathcal{Q}$ are obtained such as
\begin{align}
    \mathcal{O}&=\{i:\tilde{x}_i\in\{-1,1\},\ i\in[1:2N_t]\}\\
    \mathcal{Q}&=\{i:-1<\tilde{x}_i<1,\ i\in[1:2N_t]\}.
\end{align}  To alleviate the computational complexity of the greedy algorithm, a partial greedy algorithm is suggested based on $\tilde{\xv}_{\rm LP}$, i.e., solution of $\mathcal{P}_4$ or $\mathcal{P}_5$. 
Unlike the full greedy algorithm, the second stage of the partial greedy algorithm performs the greedy search on the entries in $\mathcal{Q}$ while elements in $\mathcal{O}$ are unchanged (see Algorithm 2) with Definition~\ref{def3},~\ref{def4}.
We can diminish the size of the search space, thereby reducing the complexity dramatically.

\subsection{Computational complexity}\label{subsec:complexity}

We compare the proposed algorithms with the existing methods in terms of the computational complexity measured by the total number of real-valued multiplications. 
We first evaluate the complexity of the optimal method based on an exhaustive search that explores all possible signal candidates $\tilde{\xv}\in\{-1,1\}^{2N_t}$. 
Since each candidate requires $2nK\cdot2N_t$ operations to generate the magnitude of coefficients in the feasibility conditions in \eqref{eq:37}, the total complexity of the exhaustive search is computed as 
\begin{equation}\label{eq:49}
    \mathcal{X}_e=4nKN_t\cdot2^{2N_t}.
\end{equation} 
We next focus on the computational complexity of the proposed algorithms which consist of LP solver of $\mathcal{P}_4$ and its greedy refinement. For the LP solver, we use the simplex method in whose computational complexity of the standard constraints of LP (i.e., $\Am\xv\ge\bv$, where $\Am\in\mathbb{R}^{m\times n}$, $\bv\in\mathbb{R}^{m\times 1},\xv\in \mathbb{R}^{n\times1}$) is given as \cite{dantzig1998linear,den2012interior}:

\begin{align}
    \mathcal{X}_{\rm LP} &= \mathbbm{i}_{\rm LP}\cdot\{(m+1)(n+1)+2 m\}\nonumber\\
    &=\mathbbm{i}_{\rm LP}\cdot\{m\times n+3m+n+1\},\label{eq:52}
\end{align}
where $\mathbbm{i}_{\rm LP}$ is the number of iterations during the simplex method. 
\textcolor{blue}{The simplex method visit all $2^{2N_t}$ vertices in worst case. Fortunately, the fact that expected complexity of the method is polynomial is proved by \cite{10.1145/990308.990310}.}
Because of the randomness of $\boldsymbol{\Lambda}$ and $\boldsymbol{\Lambda}_b$, $\mathbbm{i}_{\rm LP}$ varies according to the constraints, however, we can obtain the approximation of average $\mathbbm{i}_{\rm LP} \approx \alpha \times m$ ,where $\exp{(\alpha)} = \log_2{(2+\frac{n}{m})}$ \cite{shu1993linear}. Note that the number of iterations totally depends on the constraints of LP instead of the objective function.
Overall, the complexity of simplex method in the system is represented as
\begin{equation}
    \mathcal{X}_{\rm LP} = (\log\{\log_2{(2+\frac{n}{m})}\}\times m)\cdot(m \times n+3m+n+1). \label{eq:X_LP}
\end{equation}
Also, the quantized LP represents the algorithm that directly quantizes the solution of $\mathcal{P}_4$ or $\mathcal{P}_5$ to generate 1-bit transmit vector using {\it sign} function, i.e., $\xv_q={\rm sign}(\xv_{\rm LP})$.
Thus, the corresponding complexity is the same as the one in the LP as
\begin{equation}
    \mathcal{X}_q = \mathcal{X}_{\rm LP}.
\end{equation}
Next, the complexity of the full-greedy method is obtained as
\begin{align}\label{eq:53}
    \mathcal{X}_{\rm F-greedy}&=2\times2nK\times2N_t
\end{align}
based on the fact that the algorithm needs to sequentially search over $\tilde{x}_i\in\{-1,1\}, \forall i\in[1:2N_t]$ and each iteration requires $2nK$ operations.
Similarly, the complexity of the partial-greedy algorithm is represented as
\begin{equation} \label{eq:X_partial}
    \mathcal{X}_{\rm P-greedy}=2\times2nK\times \text{card}(\mathcal{Q}).
\end{equation}
Combining \eqref{eq:X_LP}, \eqref{eq:53}, and \eqref{eq:X_partial}, the total computational complexities of the proposed methods are computed as
\begin{align}
    \mathcal{X}_{\rm pro1}&=\mathcal{X}_{\rm LP}+\mathcal{X}_{\rm F-greedy}\nonumber\\
    &=\mathcal{X}_{\rm LP}+8nKN_t,\\
    \mathcal{X}_{\rm pro2}&=\mathcal{X}_{\rm LP}+\mathcal{X}_{\rm P-greedy}\nonumber\\
    &=\mathcal{X}_{\rm LP}+4nK\text{card}(\mathcal{Q}).\label{eq:54}
\end{align}
As a benchmark method, we consider a low-complexity method, the symbol-scaling method (SS) \cite{li2018massive} whose computational complexity is given as
\begin{equation}\label{eq:50}
    \mathcal{X}_{\rm SS}=4N_t^2+24nKN_t-2nK.
\end{equation}
\textcolor{blue}{Also, computation complexities of SQUID, C1PO and IDE \cite{wang2018finite,castaneda20171,jacobsson2016nonlinear} are given as
\begin{align}
    \mathcal{X}_{\rm SQUID} = &\mathbbm{i}_{\rm SQUID}(2 N_t K+N_t) \nonumber\\
    &+\frac{1}{3}K^3+2N_t K(K+1)+K^2,
\end{align}
\begin{align}
    \mathcal{X}_{\rm C1PO} = \mathbbm{i}_{\rm C1PO}(2 N_t K^2+N_t)+\frac{1}{3}K^3+N_t K^2,
\end{align}
\begin{align}
    \mathcal{X}_{\rm IDE} = \mathbbm{i}_{\rm IDE}(N_t K+\frac{4}{3}K^3+5N_t K+3N_t+K)\nonumber \\
    +N_t K^2,
\end{align}
where $\mathbbm{i}_{\rm x}$ denotes the number of iterations during the ${\rm x}$ algorithms, respectively. For simulation and TABLE~\ref{table:1}, we set $\mathbbm{i}_{\rm SQUID}=100,\  \mathbbm{i}_{\rm C1PO}=24,\ \mathbbm{i}_{\rm IDE}=100$, which is the setting for each paper provided \cite{wang2018finite,castaneda20171,jacobsson2016nonlinear}.}

The P-BB and OPSU algorithm are proposed for QAM constellation with low complexity in \cite{li2020interference}. Recall that the full B\&B has a prohibitive complexity in real equipment \cite{landau2017branch}. The P-BB also is based on $\tilde{x}_{\rm LP}$ and corresponding $\mathcal{O}, \mathcal{Q}$. 
In detail, P-BB fix entries of $\tilde{\xv}_{\rm LP}$, which satisfy $\tilde{x}_i\in\tilde{\xv}_{\rm LP}, i\in\mathcal{O}$, but only reconstruct the rest entries of $\tilde{\xv}_{\rm LP}$(i.e., $\tilde{x}_i\in\tilde{\xv}_{\rm LP}, i\in\mathcal{Q}$) based on B\&B. 
Moreover, the P-BB algorithm significantly reduces the complexity compared to F-BB. 
Unfortunately, in the worst case, it needs to search all subset $\{-1,1\}^{\text{card}(\mathcal{Q})}$, causing high complexity with many users. 
The OPSU algorithm is essentially a greedy algorithm based on B\&B. 
The complexity of P-BB ($\mathcal{X}_{\rm P-BB}$) cannot be specific due to B\&B search tree by case. It's totally natural that the computation complexity of OPSU ($\mathcal{X}_{\rm OPSU}$) is even lower than $\mathcal{X}_{\rm P-BB}$ \cite{li2020interference}.
The complexity of OPSU is represented as 
\begin{align}
    \mathcal{X}_{\rm OPSU} &= \mathcal{X}_{\rm LP}+2(2K+2K) \text{card}(\mathcal{Q}) \\
    &=\mathcal{X}_{\rm LP}+8K\text{card}(\mathcal{Q}).
\end{align}
In nature, we have:
\begin{equation}\label{com_pbb,opsu}
    \mathcal{X}_{\rm P-BB} \gg \mathcal{X}_{\rm OPSU}.
\end{equation} 

\textcolor{blue}{The complexity analysis in the TABLE~\ref{table:1} suppose the assumptions such as a single alternation stage, the cardinality of partial set, and the number of iterations of LP. Moreover, pivoting of the simplex method depends on the rank(\Am), which is constraint matrix of standard LP form. From 
\textit{Lemma \ref{Lemma_1}}\cite{li2020interference}, the rank$(\boldsymbol{\Lambda})$ equals rank of LP constraints for OPSU and P-BB as $2K$. Therefore, in real, there is not much difference between the complexity of OPSU and complexity of our proposed methods. Therefore, for the realistic comparison, we demonstrate the run-time simulation in Figs.~\ref{fig:runtime} and \ref{fig:runtime_detail}.}
\begin{lemma}\label{Lemma_1}
$rank(\boldsymbol{\Lambda})=2K$ with flat-fading Rayleigh fading channel $\Hm$ with probability 1.

\textit{proof}: See Appendix \ref{proof_Lemma_1}.

\end{lemma}
\section{Simulation Results}\label{sec:simulation}

\begin{figure}[!t]
    \centering
    \includegraphics[width=0.48\textwidth]{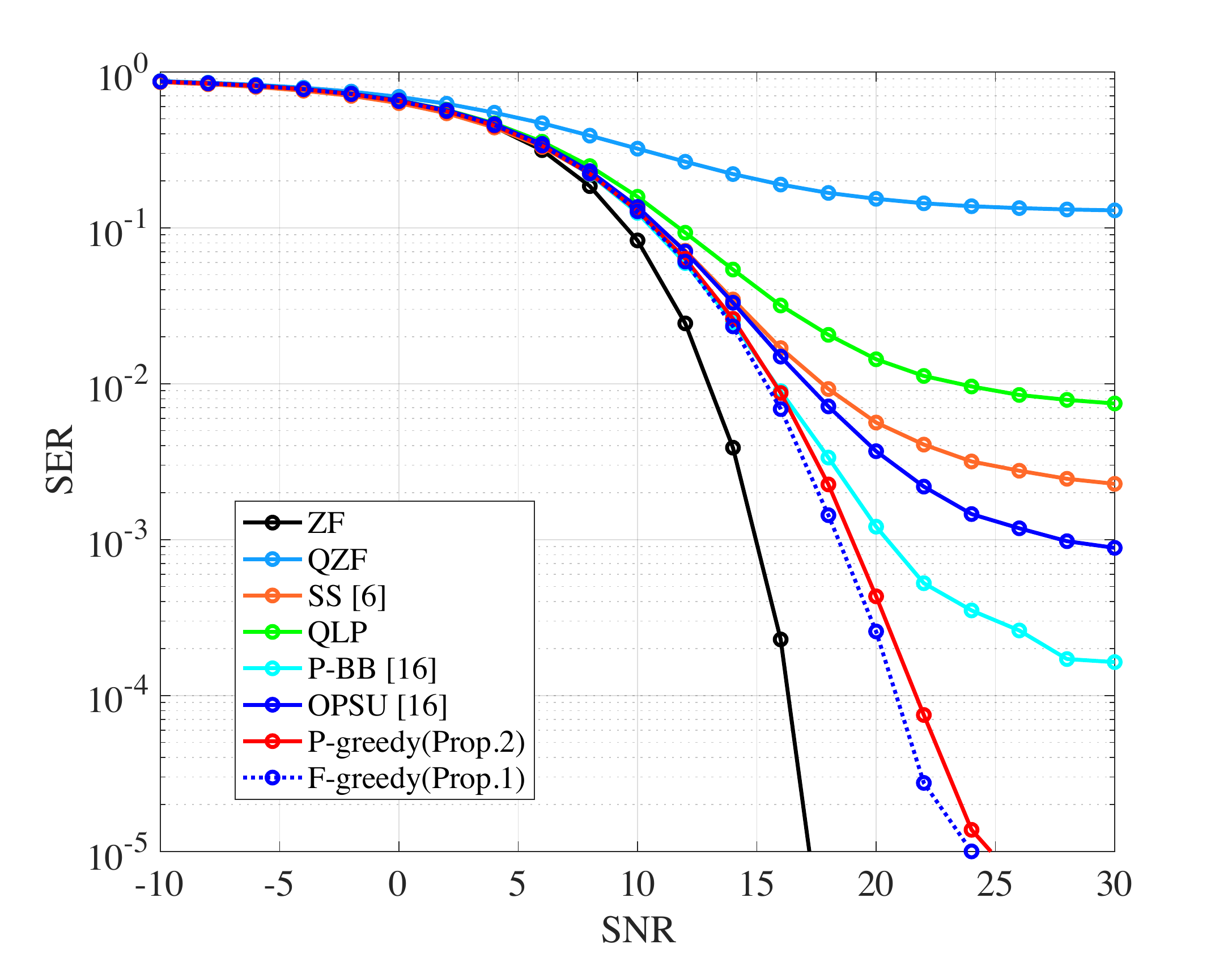}
    \caption{Performance comparisons of precoding methods for the downlink MU-MISO systems with 1-Bit DACs, where $N_t$=64, $K$=8, and $4^2$-QAM with a fixed $\tau$.}
    \label{fig:5}
\end{figure}

\begin{figure}[!t]
    \centering
    \includegraphics[width=0.48\textwidth]{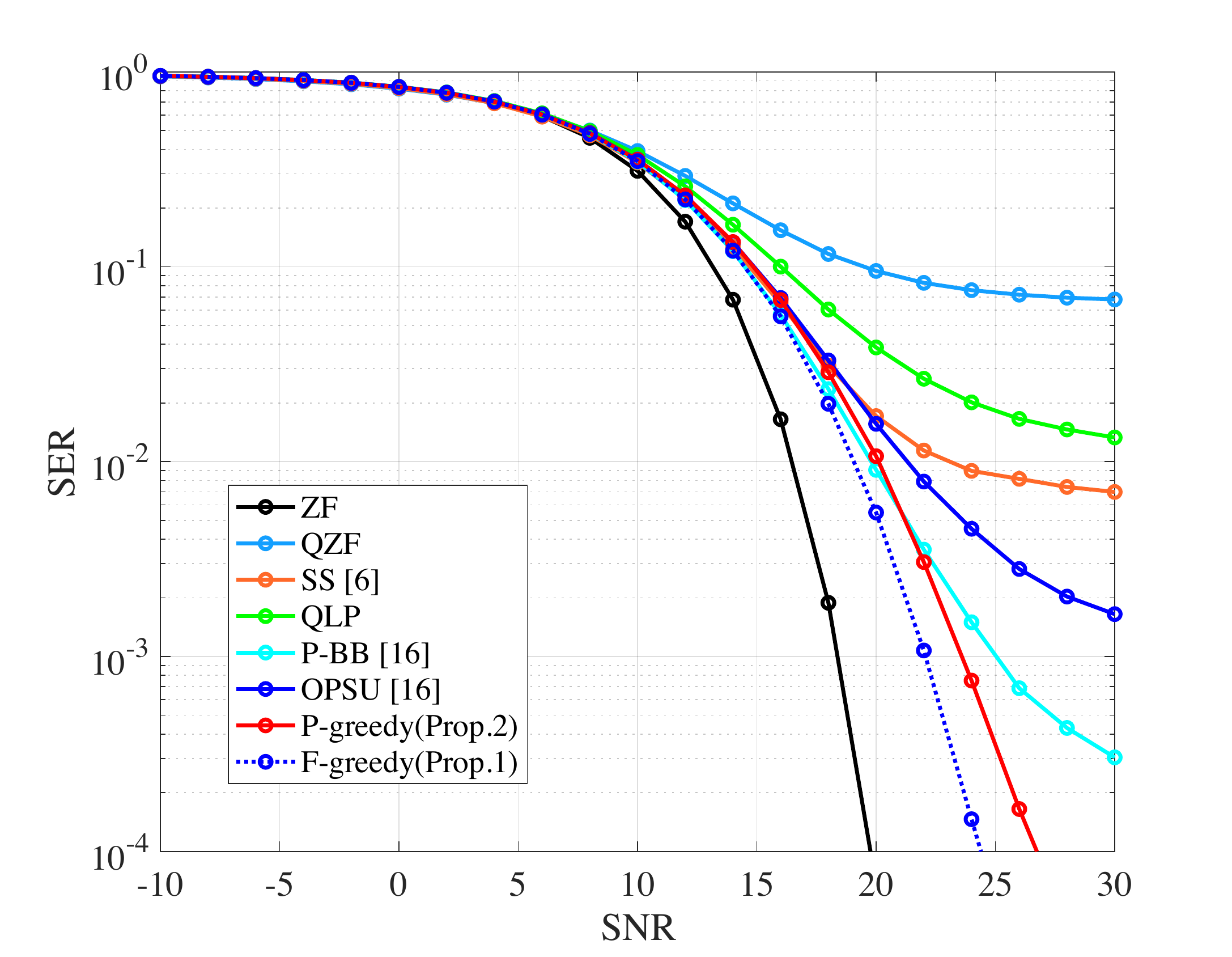}
    \caption{Performance comparisons of precoding methods for the downlink MU-MISO systems with 1-Bit DACs, where $N_t$=128, $K$=8, and $4^3$-QAM with fixed $\tau$.}
    \label{fig:6}
\end{figure}

\begin{figure}[!t]
    \centering
    \includegraphics[width=0.48\textwidth]{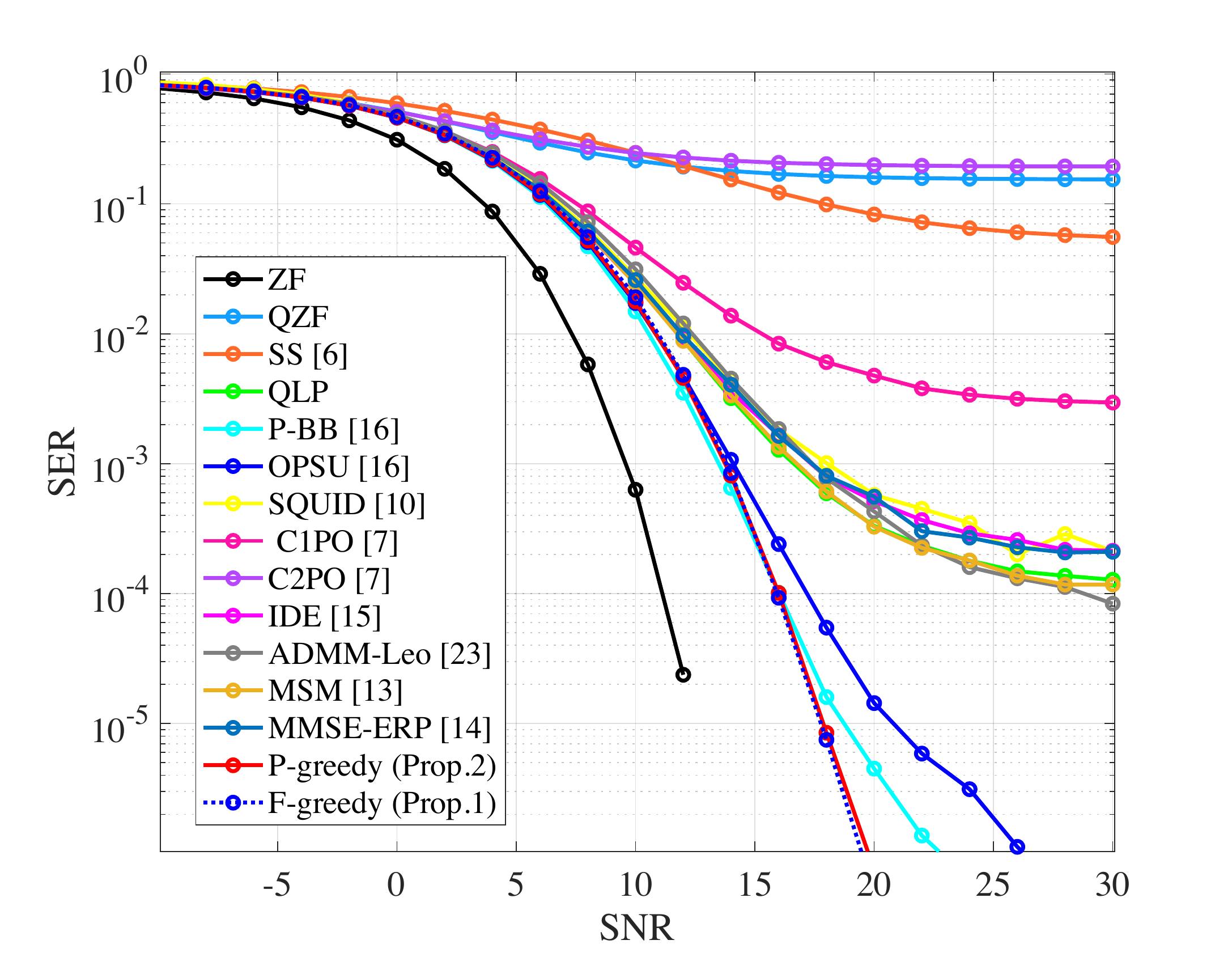}
    \caption{Performance comparisons of precoding methods for the downlink MU-MISO systems with 1-Bit DACs, where $N_t$=64, $K$=8, and $4^2$-QAM with adaptive $\tau$.}
    \label{fig:7}
\end{figure}

\begin{figure}[!t]
    \centering
    \includegraphics[width=0.48\textwidth]{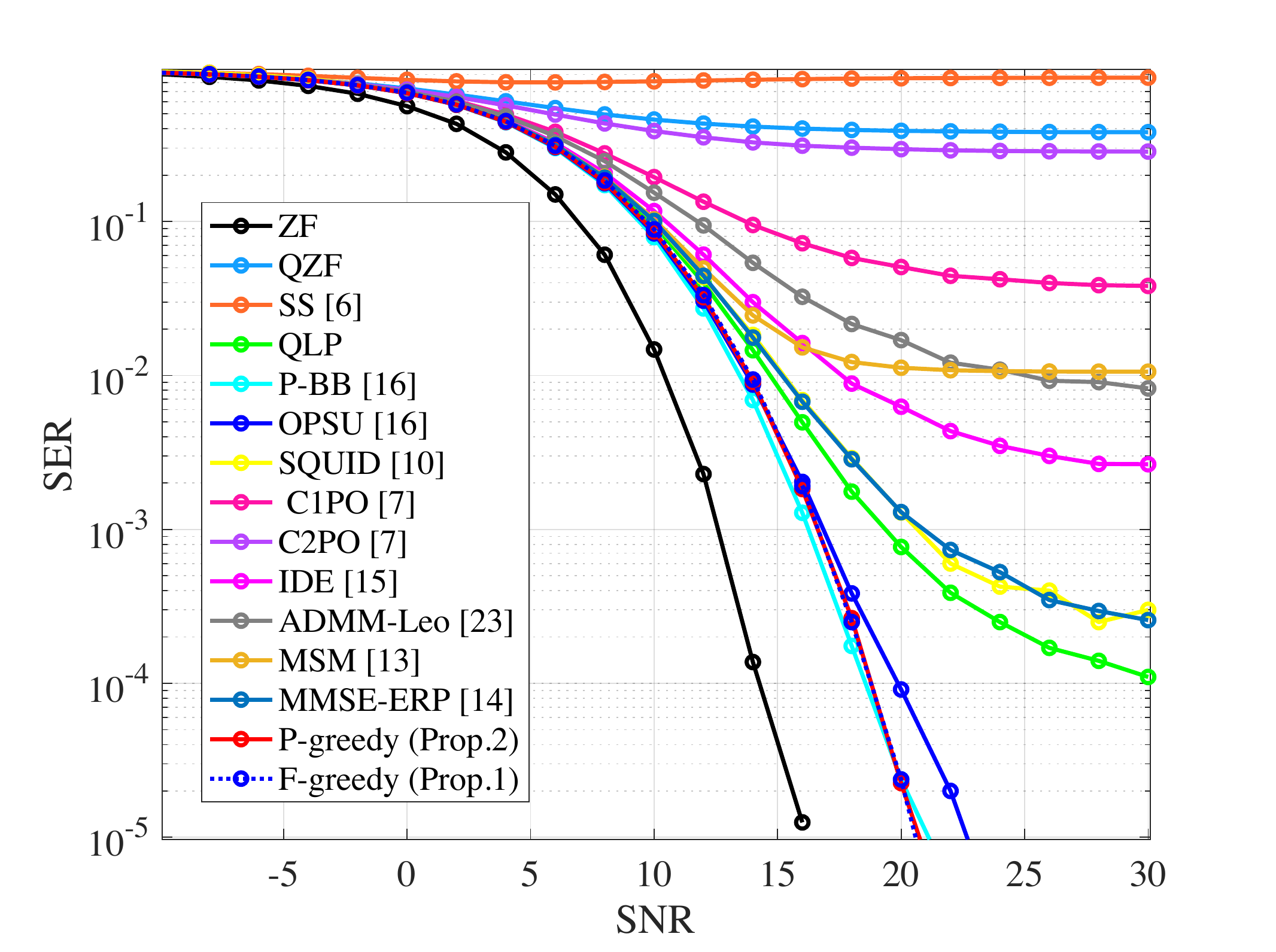}
    \caption{Performance comparisons of precoding methods for the downlink MU-MISO systems with 1-Bit DACs, where $N_t$=128, $K$=8, and $4^3$-QAM with adaptive $\tau$.}
    \label{fig:8}
\end{figure}

In this section, we verify the superiority of the proposed methods by comparing the symbol-error-rate (SER) performances of existing methods. Simulations include the following methods:

\begin{itemize}
    \item Zero forcing {\bf(ZF)}: The conventional ZF method for unquantized MU-MIMO systems, which is used as the lower-bound of the 1-bit quantized methods.
    \item Quantized zero forcing {\bf (QZF)}: The direct 1-bit  quantization of ZF.
    \item Symbol scaling {\bf (SS)}: The low complexity method proposed in \cite{li2018massive}.
    \item  Quantized LP {\bf (QLP)}: The direct 1-bit quantization of the solution from $\mathcal{P}_4$ in Figs. \ref{fig:7}, \ref{fig:8} and $\mathcal{P}_5$ in Figs.  \ref{fig:5}, \ref{fig:6}.
    \item Partial branch and bound {\bf (P-BB)}: The method for QAM constellations proposed in \cite{li2020interference}. 
    \item Ordered partial sequential update {\bf (OPSU)}: The ordered greedy method based on B\&B proposed in \cite{li2020interference}.
  \textcolor{blue}{  \item Squared-infinity norm Douglas-Rachford splitting {\bf (SQUID)}: The well-known algorithm that have excellent performance in \cite{jacobsson2016nonlinear}.
    \item Biconvex 1-bit precoding {\bf (C1PO, C2PO)}: The low complexity algorithms for any constellations in \cite{castaneda20171}.
    \item Iterative discrete estimation {\bf (IDE)}: The iteration algorithm based on ADMM in \cite{wang2018finite}.
    \item ADMM-Leo {\bf (ADMM-Leo)}: The efficient algorithm based on ADMM in \cite{chu2019efficient}.
    \item MSM method {\bf (MSM)}: The constant envelope precoding for MSM design criterion in \cite{jedda2018quantized}.    
    \item MMSE with enhanced receive processing {\bf (MMSE-ERP)}: The MMSE based precoding using alternating minimization in \cite{chen2019mmse}.}
    \item Full-greedy {\bf (F-greedy)}: The proposed method 1, full greedy algorithm based LP from Algorithm 1.
    \item Partial-greedy {\bf (P-greedy)}: The proposed method 2, partial greedy algorithm based LP from Algorithm 2.

\end{itemize}
Recall that $\SNR$ is defined as per-antenna signal-to-noise ratio, i.e., $\rho/\sigma^2$.  
\begin{table*}[h]
\caption{Comparison of computation complexities at setting of performances}
\begin{center}
\begin{tabular}{|c|c|c|}
\hline
Precoding methods & $4^2$-QAM, $N_t=64, K=8$& $4^3$-QAM, $N_t=128, K=8$\\
\hline 
Exhaustive search&  $1.4\times10^{42}$ &$1.4\times10^{81}$  \\
\cline{1-3}
P-BB& $69481\ll\cdot\leq1.34\cdot10^8$ &$263030\ll\cdot\leq4\cdot10^8$ \\
\hline
\cline{1-3}
OPSU& 69481 & 263030 \\
\hline
\cline{1-3}
QLP& 131320 & 643100 \\
\hline
\cline{1-3} 
F-greedy& 139510 &667680 \\
\hline
\cline{1-3}
P-greedy& 133240 &645980 \\
\hline
\cline{1-3}
\cline{1-3}
SS& 40928 &139216 \\
\hline
\cline{1-3}
SQUID (iteration=100)& 118250 &236270 \\
\hline
\cline{1-3}
C1PO (iteration=24)& 31915 &62123 \\
\hline
\cline{1-3}
IDE (iteration=100)& 757960 &1446900 \\
\hline

\end{tabular}
\label{table:1}
\end{center}
\end{table*}
In addition, we evaluate the computation complexities based on the analyses in Section \ref{subsec:complexity}. 
For complexity of OPSU and P-greedy algorithms in TABLE~\ref{table:1}, We also use average $\text{card}(Q)$ from $10^5$ simulations, and 
the average $\text{card}(Q)$ in the setting ($4^2$-QAM, $N_t=64$, $K=8$) is $14.988$ 
Also, in the setting of ($4^3$-QAM, $N_t=128$, $K=8$), the average $\text{card}(Q)$ is $14.998$.
TABLE~\ref{table:1} shows that the computation complexity of proposed methods are moderate \cite{li20201bit}. 

The following two scenarios are considered according to the overhead for information on a decision-region $\tau$:
\begin{itemize}
\item[] \textbf{Scenario i)} $\tau$ is chosen regardless of a channel matrix $\Hm$, i.e., $\tau$ is determined by (\ref{eq:10}). The corresponding results are provided in Figs. \ref{fig:5} and \ref{fig:6}. The $\tilde{\xv}_{\rm LP}$ and $\tau$, inputs of our proposed methods (i.e., F-greedy and P-greedy algorithms) is determined from $\mathcal{P}_5$. \\
\item[] \textbf{Scenario ii)} Figs. \ref{fig:7} and  \ref{fig:8} adopt $\tau$ from own schemes. In scenario ii), the solutions of $\mathcal{P}_4$, $\tilde{\xv}_{\rm LP}$ and $\tau(\eqdef t)$ are equipped at the proposed algorithms. Although we have to transmit $\tau$ depending on channel $\Hm$ at given time, the performance gain is attractive.
\end{itemize}

Fig.~\ref{fig:5} shows the SER performance comparisons of the above algorithms for downlink MU-MISO systems with 1-bit DACs where $N_t=64$, $K=8$, and $4^2$-QAM. 
Without 1-bit constraints, the ZF method provides an optimal performance with infinite-resolution DACs.
This can be interpreted as the lower-bound of the above 1-bit constraint methods. Note that, in all simulation settings, we cannot evaluate the performance of MILP due to its unmanageable complexity.
At high SNR, we observe that all 1-bit precoding methods including the quantized LP suffer from a severe error-floor except the proposed methods.
To overcome the error-floor, we apply the F-greedy and P-greedy algorithms which determine the entries of $\xv_{\rm LP}$ such that they belong to $\{-1,1\}$ while keeping the feasibility and robustness. 

Fig.~\ref{fig:6} shows the SER performance comparisons for the configuration of $N_t=128$, $K=8$, and $4^3$-QAM showing the similar trend. Based on the formulation of the optimization problem, our formulation has all candidates in the decision region due to the expression of intersections of the $n$ shifted base regions. In detail, it causes that our MILP, LP problem can have only inequality constraint without equality constraint. 

Fig.~\ref{fig:7} shows the performance comparisons of the MU-MISO case where $N_t=64$, $K=8$ and $4^2$-QAM with adaptive $\tau$. 
We observe the performances of the proposed methods are the closest to the optimal performance, however the deviation between the full greedy method and the partial greedy method is trivial, which means the $\mathcal{P}_4$ provides near optimal $\tau$ and the refinement of $\xv_{\rm{LP}}$ over $\mathcal{Q}$ is sufficient. 
At high SNR, the proposed methods show more performance gain over P-BB which is the near optimal method \cite{li2020interference}, which further validates that $\tau$ from $\mathcal{P}_4$ is properly chosen.

Fig.~\ref{fig:8} also shows the same aspect of the systems, where $N_t=128$, $K=8$ and $4^3$-QAM with adaptive $\tau$. Although we assume that the iteration is only once in the actual algorithm process, P-BB and OPSU find $\tau$ alternatively, whereas the proposed method fix the $\tau$ from $\mathcal{P}_4$ at a sitting. The performances in Figs.~\ref{fig:7} and \ref{fig:8} show that $\tau$ found at once in $\mathcal{P}_4$ is reasonable.

\begin{figure}
    \centering
    \includegraphics[width=0.48\textwidth]{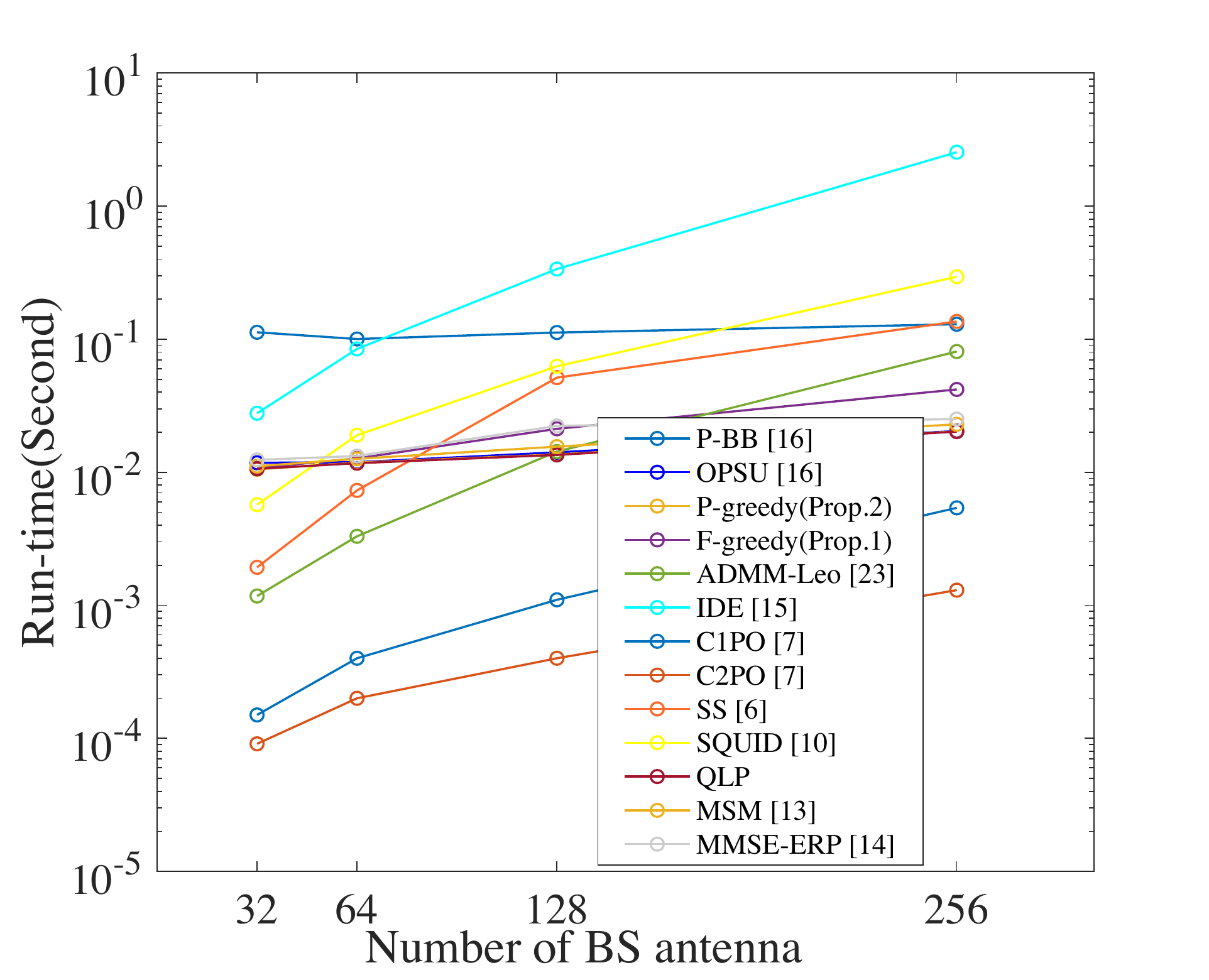}
    \caption{Run-time versus the number of BS antennas for precoding methods , where $K$=8, and $4^2$-QAM with adaptive $\tau$.}
    \label{fig:runtime}
\end{figure}

\begin{figure}
    \centering
    \includegraphics[width=0.48\textwidth]{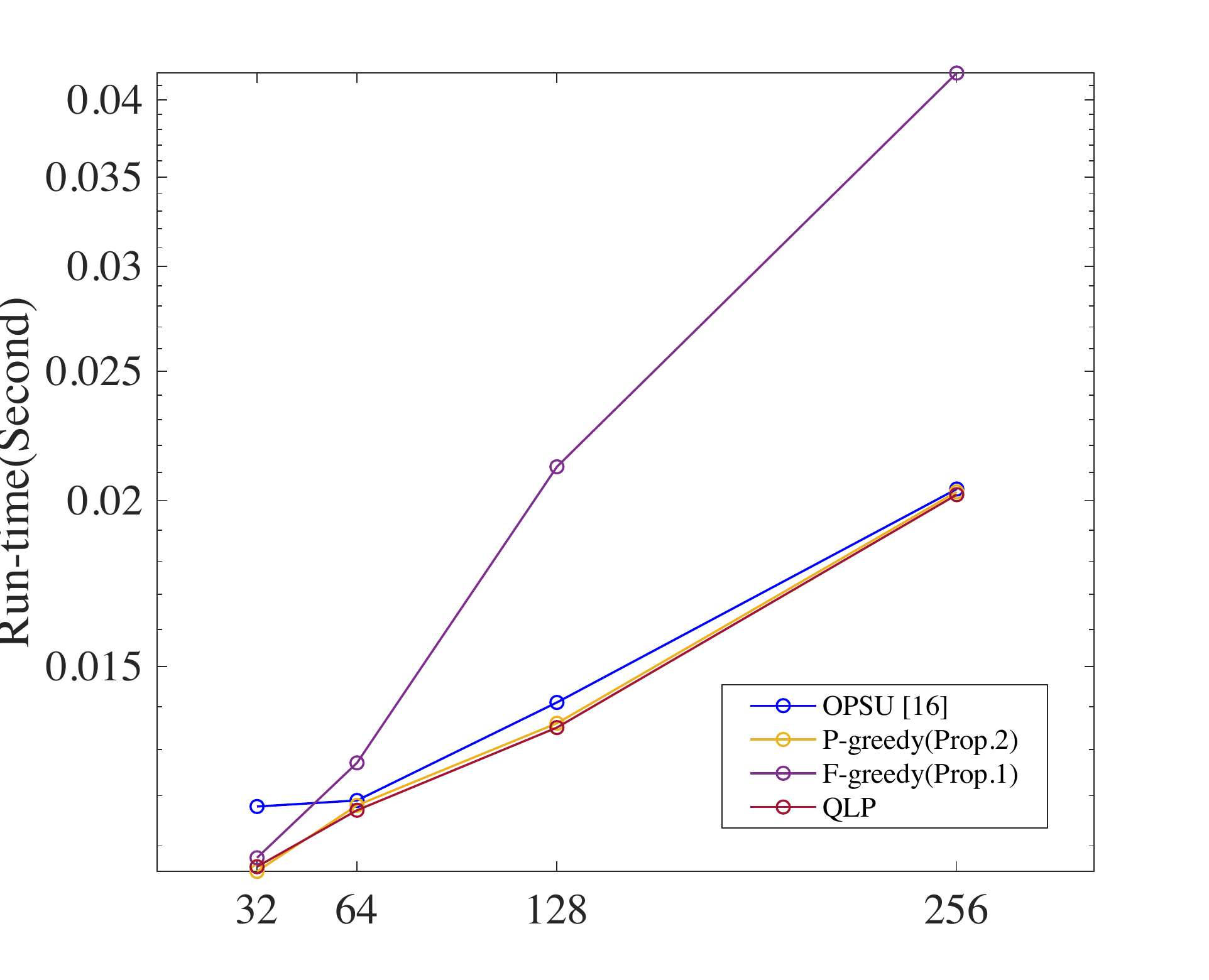}
    \caption{Run-time versus the number of BS antennas for QLP, OPSU, F-greedy, P-greedy algorithms, where $K$=8, and $4^2$-QAM with adaptive $\tau$.}
    \label{fig:runtime_detail}
\end{figure}

\begin{figure}[!t]
    \centering
    \includegraphics[width=0.48\textwidth]{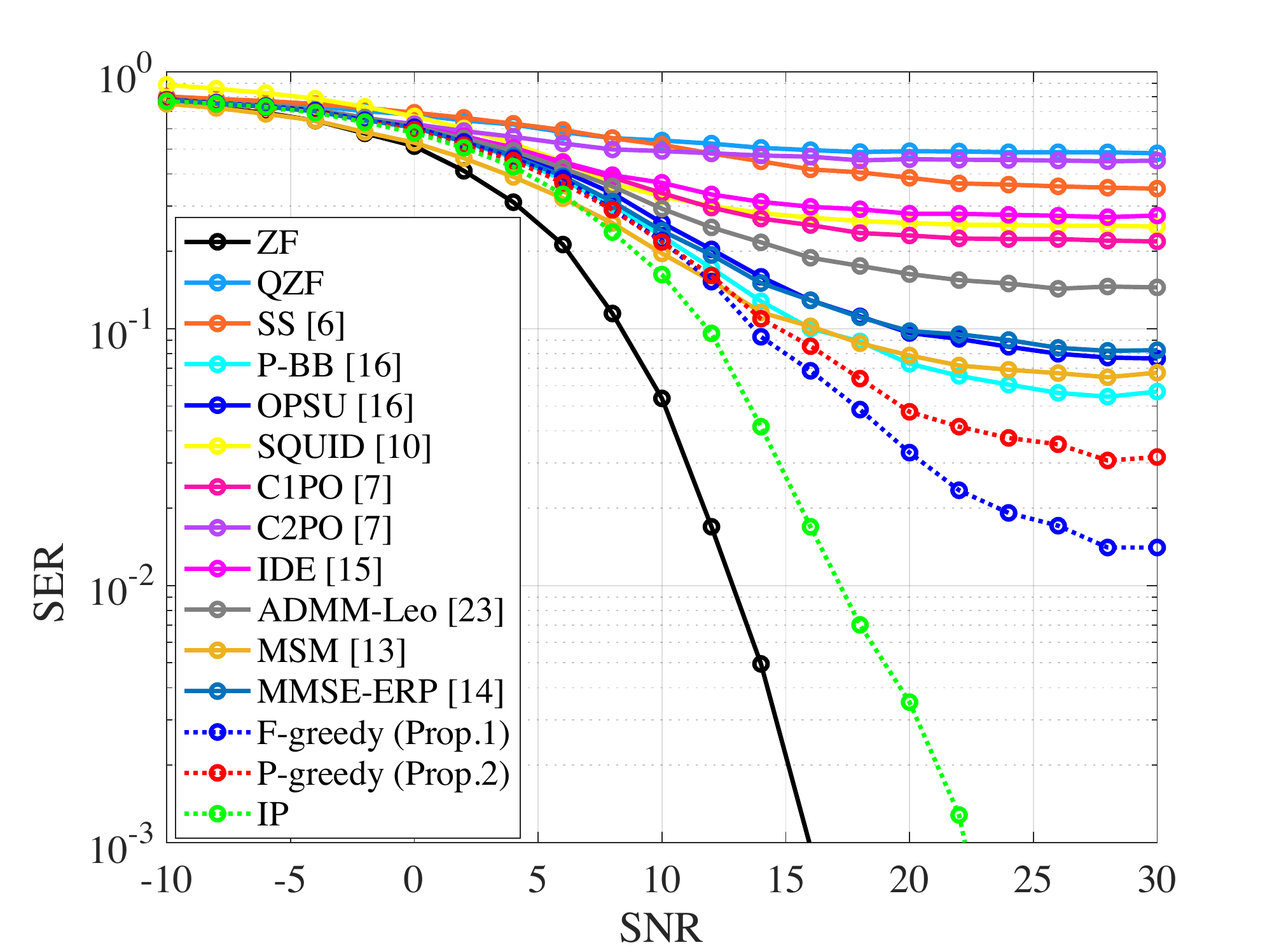}
    \caption{Performance comparisons of precoding methods for the downlink MU-MISO systems with 1-Bit DACs, where $N_t$=8, $K$=2, and $4^2$-QAM with adaptive $\tau$.}
    \label{fig:smallscale}
\end{figure}

\textcolor{blue}{Fig.~\ref{fig:smallscale} shows the performance of IP, solution from $\mathcal{P}_3$ in small-scale MIMO. The other algorithms perform undesirable performance. However, the proposed methods achieve some suitable performance and IP attains near-optimal performance. 
Unfortunately, high complexity of $\mathcal{P}_4$ mainly caused by large-scale antenna array prevents the simulation. The performance loss of IP is caused by the adaptive $\tau$. $\tau$ of IP is generally smaller than $\tau$ of ZF with infinite resolution due to difference of candidates by resolutions. We already provide computational complexity of many methods in \ref{subsec:complexity}. However, many assumption of the computations are existed. For the specific comparison, we compare run-time of the methods in same computer setting with $10^4$ simulations. Fig.~\ref{fig:runtime} and Fig.~\ref{fig:runtime_detail} shows the proposed methods have short run-times. In TABLE~\ref{table:1}, the complexity of OPSU is almost half of the F-greedy. However, in real, the complexity of LP is almost same.
In addition, the run-time of LP mainly depends on the number of users unlike other methods. we expect to compensate the performance loss from resolution with more BS antennas without complexity loss.}

\begin{figure}[!t]
    \centering
    \includegraphics[width=0.48\textwidth]{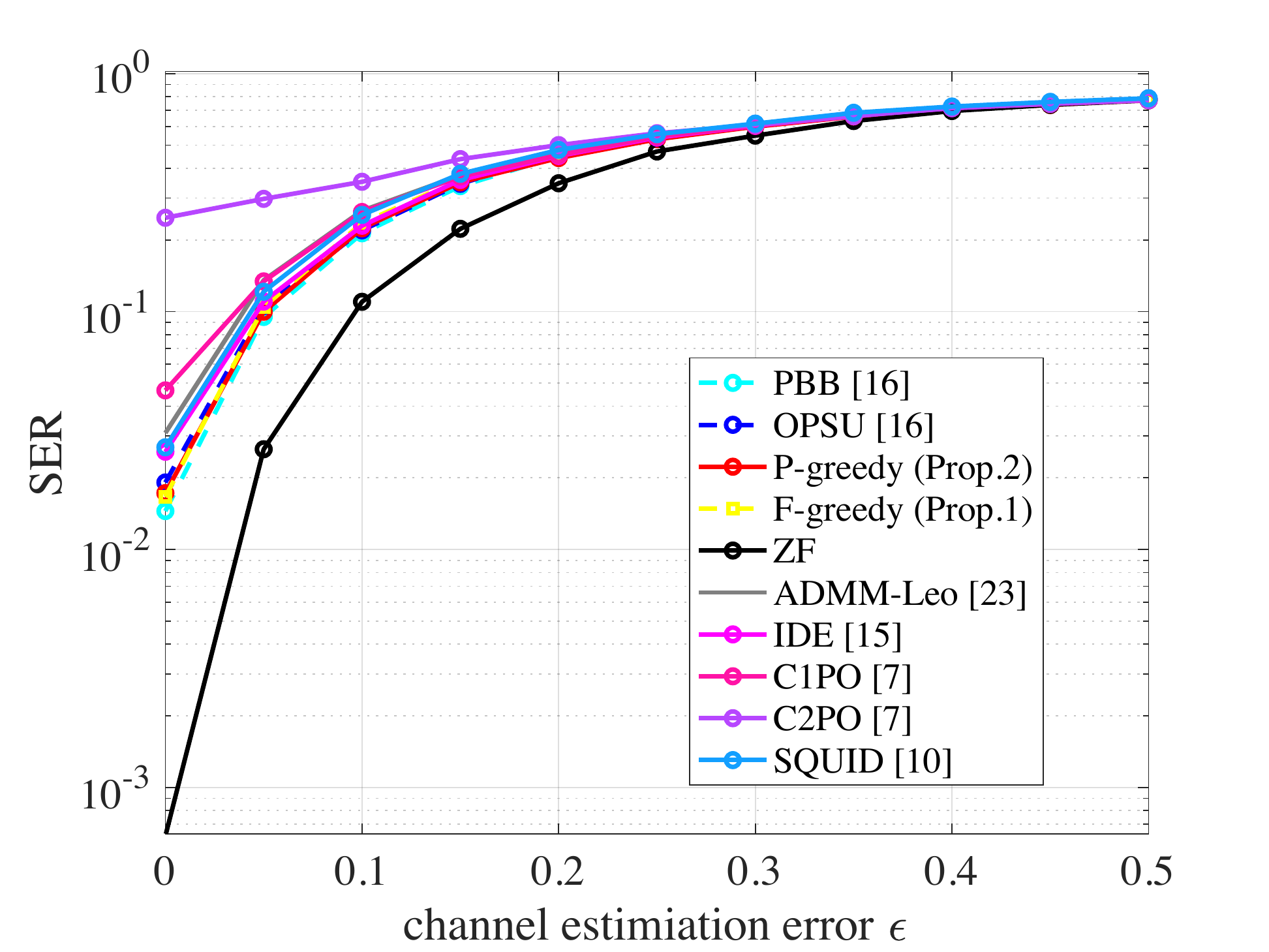}
    \caption{Performance comparisons of precoding methods for the downlink MU-MISO systems with 1-Bit DACs for channel estimiation error $\epsilon$, where $N_t$=64, $K$=8, and $4^2$-QAM with adaptive $\tau$ in $SNR$=10.}
    \label{fig:channel_error}
\end{figure}

\textcolor{blue}{Fig.~\ref{fig:runtime} shows novelty of our algorithms in terms of the computation complexity. The run-time of each algorithms is the averaged over $10^4$ simulations. Run-time of our algorithms are attractive when having large-scale antennas arrays, i.e., massive MIMO systems. From performance figures and Fig.~\ref{fig:runtime}, the proposed algorithms perform the same as P-BB, the near-optimal performance. But, run-time is about 10 times less. Furthermore, we exploit the simplex method where the complexity almost depend on the number of users. Therefore the run-times of the methods based on LP rarely increase. So, we can consider the scenario that cover up the performance loss as more antennas at BS instead of increasing power of all transmit antennas. }

\textcolor{blue}{
We further present the robustness of our algorithms to channel estimation errors. At BS, we assume the imperfect CSI as
\begin{equation}
    \Hm_{e} = \sqrt{1-\epsilon}\Hm + \sqrt{\epsilon}\Em,
\end{equation}
where $\epsilon\in[0,1]$ and $\Em\in\mathbbm{C}^{K\times N_t}$. Therefore, $\epsilon=0, \epsilon\in(0,1)$, and $\epsilon=1$ mean perfect CSI, partial CSI and a no CSI scenario, respectively. In fig.~\ref{fig:channel_error}, the proposed algorithms still achieve near-optimal performance with 10 {\rm dB} SNR under the imperfect CSI.  
}

\section{Conclusion}\label{sec:conclusion}
We proposed the construction of 1-bit transmit signal vector for downlink MU-MISO systems with QAM constellations. In this regard, we derived the linear feasibility constraints which ensure that each user can recover the desired message successfully and transformed them into the cascaded matrix form. From this, we constructed mixed integer linear programming (MILP) problem whose solution generates a 1-bit transmit vector to satisfy the feasibility conditions and guarantee the robustness to a noise. To address the computational complexity of MILP, we proposed the LP-relaxed algorithm consisting of two steps: 1) to solve the relaxed LP; 2) to refine the LP solution to fit into the 1-bit constraint. Via simulation results, we demonstrated that the proposed methods show better performances with low-complexity compared with the benchmarks. One promising future direction is to further reduce the complexity of the proposed method without the performance loss. 
 \begin{appendices}
\section{Proof for Lemma 1}\label{proof_Lemma_1}
$\mathcal{W}$ denotes the space spanned by rows of $\bar{\Hm}$. Any $\wv^{\transp}\in\mathcal{W}$ is represented as a linear combination of the row of $\bar{\Hm}$
\begin{equation}\label{eq_appA:1}
    \wv^{\transp} = \vv^{\transp}\bar{\Hm},
\end{equation}
where $\vv^{\transp}$ is $1\times2nK$ vector.
$\bar{\Mm}$ is full rank because it is diagonal matrix by (\ref{eq:28}). It has $2nK$ linear independent rows which span the space of $2nK$ dimension. Therefore, There exists a $1\times2nK$ vector $\uv^{\transp}$ such that 
\begin{equation}\label{eq_appA:2}
    \vv^{\transp} = \uv^{\transp}\bar{\Mm}.    
\end{equation}
And then, by (\ref{eq_appA:1}), (\ref{eq_appA:2}), $\wv^{\transp}$ is represented as
\begin{equation}
    \wv^{\transp} = \vv^{\transp}\bar{\Hm} = (\uv^{\transp}\bar{\Mm})\bar{\Hm} = \uv^{\transp}(\bar{\Mm}\bar{\Hm})
\end{equation}
In detail, $\wv^{\transp}$ is a linear combination of the rows of $\bar{\Mm\bar{\Hm}}$ and a linear combination of rows of $\bar{\Hm}$ as well.  
By definition of rank,
\begin{equation}
    \text{rank}(\boldsymbol{\Lambda}) = \text{rank}(\bar{\Mm}\bar{\Hm}) = \text{rank}(\bar{\Hm}).
\end{equation}
$\Hm$ is flat-fading Rayleigh fading channel that has full rank and real part and imaginary part of $\Hm$ are i.i.d..
We could see $\text{rank}(\bar{\Hm}) = 2K$ based on (\ref{eq:26}), (\ref{eq:32}) and $\text{rank}(\tilde{\Hm}) = 2K$.
Overall, proof of Lemma 1\ref{Lemma_1} is completed as rank of $\boldsymbol{\Lambda}$ is 2K.
\end{appendices}

\section*{Acknowledgment}
This work was supported by the National Research Foundation of Korea (NRF) grant funded by the Korea government (MSIT) (NRF-2020R1A2C1099836).



\bibliographystyle{IEEEtran}
\bibliography{journal_1bit_v3.bib}

\begin{thebibliography}{10}
\providecommand{\url}[1]{#1}
\csname url@samestyle\endcsname
\providecommand{\newblock}{\relax}
\providecommand{\bibinfo}[2]{#2}
\providecommand{\BIBentrySTDinterwordspacing}{\spaceskip=0pt\relax}
\providecommand{\BIBentryALTinterwordstretchfactor}{4}
\providecommand{\BIBentryALTinterwordspacing}{\spaceskip=\fontdimen2\font plus
\BIBentryALTinterwordstretchfactor\fontdimen3\font minus
  \fontdimen4\font\relax}
\providecommand{\BIBforeignlanguage}[2]{{%
\expandafter\ifx\csname l@#1\endcsname\relax
\typeout{** WARNING: IEEEtran.bst: No hyphenation pattern has been}%
\typeout{** loaded for the language `#1'. Using the pattern for}%
\typeout{** the default language instead.}%
\else
\language=\csname l@#1\endcsname
\fi
#2}}
\providecommand{\BIBdecl}{\relax}
\BIBdecl

\bibitem{marzetta2010noncooperative}
T.~L. {Marzetta}, ``{Noncooperative Cellular Wireless with Unlimited Numbers of
  Base Station Antennas},'' \emph{IEEE Trans. on Wireless Commun.}, vol.~9,
  no.~11, pp. 3590--3600, 2010.

\bibitem{spencer2004introduction}
Q.~H. Spencer, C.~B. Peel, A.~L. Swindlehurst, and M.~Haardt, ``An introduction
  to the multi-user mimo downlink,'' \emph{IEEE communications Magazine},
  vol.~42, no.~10, pp. 60--67, 2004.

\bibitem{larsson2014massive}
E.~G. Larsson, O.~Edfors, F.~Tufvesson, and T.~L. Marzetta, ``Massive mimo for
  next generation wireless systems,'' \emph{IEEE communications magazine},
  vol.~52, no.~2, pp. 186--195, 2014.

\bibitem{peel2005vector}
C.~B. Peel, B.~M. Hochwald, and A.~L. Swindlehurst, ``A vector-perturbation
  technique for near-capacity multiantenna multiuser communication-part i:
  channel inversion and regularization,'' \emph{IEEE Transactions on
  Communications}, vol.~53, no.~1, pp. 195--202, 2005.

\bibitem{park2019construction}
G.-J. Park and S.-N. Hong, ``Construction of 1-bit transmit-signal vectors for
  downlink mu-miso systems with psk signaling,'' \emph{IEEE Transactions on
  Vehicular Technology}, vol.~68, no.~8, pp. 8270--8274, 2019.

\bibitem{li2018massive}
A.~Li, C.~Masouros, F.~Liu, and A.~L. Swindlehurst, ``Massive mimo 1-bit dac
  transmission: A low-complexity symbol scaling approach,'' \emph{IEEE
  Transactions on Wireless Communications}, vol.~17, no.~11, pp. 7559--7575,
  2018.

\bibitem{castaneda20171}
O.~Casta{\~n}eda, S.~Jacobsson, G.~Durisi, M.~Coldrey, T.~Goldstein, and
  C.~Studer, ``1-bit massive mu-mimo precoding in vlsi,'' \emph{IEEE Journal on
  Emerging and Selected Topics in Circuits and Systems}, vol.~7, no.~4, pp.
  508--522, 2017.

\bibitem{landau2017branch}
L.~T. Landau and R.~C. de~Lamare, ``Branch-and-bound precoding for multiuser
  mimo systems with 1-bit quantization,'' \emph{IEEE Wireless Communications
  Letters}, vol.~6, no.~6, pp. 770--773, 2017.

\bibitem{amor201716}
D.~B. Amor, H.~Jedda, and J.~Nossek, ``16 qam communication with 1-bit
  transmitters,'' in \emph{WSA 2017; 21th International ITG Workshop on Smart
  Antennas}.\hskip 1em plus 0.5em minus 0.4em\relax VDE, 2017, pp. 1--5.

\bibitem{jacobsson2016nonlinear}
S.~Jacobsson, G.~Durisi, M.~Coldrey, T.~Goldstein, and C.~Studer, ``Nonlinear
  1-bit precoding for massive mu-mimo with higher-order modulation,'' in
  \emph{2016 50th Asilomar Conference on Signals, Systems and Computers}.\hskip
  1em plus 0.5em minus 0.4em\relax IEEE, 2016, pp. 763--767.

\bibitem{sohrabi2018one}
F.~Sohrabi, Y.-F. Liu, and W.~Yu, ``One-bit precoding and constellation range
  design for massive mimo with qam signaling,'' \emph{IEEE Journal of Selected
  Topics in Signal Processing}, vol.~12, no.~3, pp. 557--570, 2018.

\bibitem{jacobsson2017quantized}
S.~Jacobsson, G.~Durisi, M.~Coldrey, T.~Goldstein, and C.~Studer, ``Quantized
  precoding for massive mu-mimo,'' \emph{IEEE Transactions on Communications},
  vol.~65, no.~11, pp. 4670--4684, 2017.

\bibitem{jedda2018quantized}
H.~Jedda, A.~Mezghani, A.~L. Swindlehurst, and J.~A. Nossek, ``Quantized
  constant envelope precoding with psk and qam signaling,'' \emph{IEEE
  Transactions on Wireless Communications}, vol.~17, no.~12, pp. 8022--8034,
  2018.

\bibitem{chen2019mmse}
C.-E. Chen, ``Mmse one-bit precoding for mu-mimo systems with enhanced receive
  processing,'' \emph{IEEE Wireless Communications Letters}, vol.~9, no.~4, pp.
  548--552, 2019.

\bibitem{wang2018finite}
C.-J. Wang, C.-K. Wen, S.~Jin, and S.-H. Tsai, ``Finite-alphabet precoding for
  massive mu-mimo with low-resolution dacs,'' \emph{IEEE Transactions on
  Wireless Communications}, vol.~17, no.~7, pp. 4706--4720, 2018.

\bibitem{li2020interference}
A.~Li, F.~Liu, C.~Masouros, Y.~Li, and B.~Vucetic, ``Interference exploitation
  1-bit massive mimo precoding: A partial branch-and-bound solution with
  near-optimal performance,'' \emph{IEEE Transactions on Wireless
  Communications}, vol.~19, no.~5, pp. 3474--3489, 2020.

\bibitem{li20201bit}
A.~Li, C.~Masouros, A.~L. Swindlehurst, and W.~Yu, ``1-bit massive mimo
  transmission: Embracing interference with symbol-level precoding,'' 2020.

\bibitem{luenberger1984linear}
D.~G. Luenberger, Y.~Ye \emph{et~al.}, \emph{Linear and nonlinear
  programming}.\hskip 1em plus 0.5em minus 0.4em\relax Springer, 1984, vol.~2.

\bibitem{den2012interior}
D.~Den~Hertog, \emph{Interior point approach to linear, quadratic and convex
  programming: algorithms and complexity}.\hskip 1em plus 0.5em minus
  0.4em\relax Springer Science \& Business Media, 2012, vol. 277.

\bibitem{dantzig1998linear}
G.~B. Dantzig, \emph{Linear programming and extensions}.\hskip 1em plus 0.5em
  minus 0.4em\relax Princeton university press, 1998, vol.~48.

\bibitem{10.1145/990308.990310}
\BIBentryALTinterwordspacing
D.~A. Spielman and S.-H. Teng, ``Smoothed analysis of algorithms: Why the
  simplex algorithm usually takes polynomial time,'' \emph{J. ACM}, vol.~51,
  no.~3, p. 385–463, May 2004. [Online]. Available:
  \url{https://doi.org/10.1145/990308.990310}
\BIBentrySTDinterwordspacing

\bibitem{shu1993linear}
F.~Shu-Cherng and S.~Puthenpura, ``Linear optimization and extensions. theory
  and algorithms,'' 1993.

\bibitem{chu2019efficient}
L.~Chu, F.~Wen, L.~Li, and R.~Qiu, ``Efficient nonlinear precoding for massive
  mimo downlink systems with 1-bit dacs,'' \emph{IEEE Transactions on Wireless
  Communications}, vol.~18, no.~9, pp. 4213--4224, 2019.

\end{thebibliography}

\end{document}